\documentclass[12pt]{interact}

\usepackage{epstopdf}
\usepackage{subfigure}
\usepackage[numbers,merge]{natbib}
\bibpunct[, ]{(}{)}{,}{n}{,}{,}
\usepackage{natbib}
\theoremstyle{plain}

\theoremstyle{definition}

\usepackage{float}
\theoremstyle{remark}

\setcitestyle{authoryear}
\usepackage{threeparttable}
\usepackage{caption}
\usepackage{hyperref}
\usepackage{url}
\usepackage{rotating}
\hypersetup{final}

\title{A Neural-embedded Discrete Choice Model:\\
Learning Taste Representation with Strengthened Interpretability
}

\begin{document}

\author{
\name{Yafei Han\textsuperscript{a},
Francisco Camara Pereira\textsuperscript{b}, 
Moshe Ben-Akiva\textsuperscript{c},
Christopher Zegras\textsuperscript{d}}
\affil{\textsuperscript{a} Civil and Environmental Engineering, Massachusetts Institute of Technology (yafei@alum.mit.edu, corresponding author)
\newline
\textsuperscript{b} Department of Technology, Management and Economics, Technical University of Denmark (camara@dtu.dk)
\newline 
\textsuperscript{c} Civil and Environmental Engineering, Massachusetts Institute of Technology (mba@mit.edu)
\newline
\textsuperscript{d} Department of Urban Studies and Planning, Massachusetts Institute of Technology (czegras@mit.edu)}
}

\maketitle

\newpage
\begin{abstract}
Discrete choice models (DCMs) require a \textit{priori} knowledge of the utility functions, especially how tastes vary across individuals. Utility misspecification may lead to biased estimates, inaccurate interpretations and limited predictability. In this paper, we utilize a neural network to learn \textit{taste representation}. Our formulation consists of two modules: a neural network (TasteNet) that learns taste parameters (e.g., time coefficient) as flexible functions of individual characteristics; and a multinomial logit (MNL) model with utility functions defined with expert knowledge. Taste parameters learned by the neural network are fed into the choice model and link the two modules. 

Our approach extends the L-MNL model \citep{SIFRINGER2020236} by allowing the neural network to learn the interactions between individual characteristics and alternative attributes. Moreover, we formalize and strengthen the interpretability condition - requiring realistic estimates of behavior indicators (e.g., value-of-time, elasticity) at the disaggregated level, which is crucial for a model to be suitable for scenario analysis and policy decisions. Through a unique network architecture and parameter transformation, we incorporate prior knowledge and guide the neural network to output realistic behavior indicators at the disaggregated level. 

We show that TasteNet-MNL reaches the ground-truth model's predictability and recovers the nonlinear taste functions on synthetic data. Its estimated value-of-time and choice elasticities at the individual level are close to the ground truth. In contrast, exemplary logit models with misspecified systematic utility lead to biased parameter estimates and lower prediction accuracy. On a publicly available Swissmetro dataset, TasteNet-MNL outperforms benchmarking MNLs and Mixed Logit model's predictability. It learns a broader spectrum of taste variations within the population and suggests a higher average value-of-time. Our source code is available for research and application. 
\end{abstract}


\newpage
\section{Introduction}
Discrete choice model (DCM) is a robust econometric framework to understand and predict choice behaviors. With a solid foundation in random utility theory, DCM has the significant advantage of interpretability: it can explain how individuals choose among a set of alternatives and provide reliable answers to ``what-if'' scenario questions, which is crucial for making policy decisions. 

A DCM requires making assumptions about a choice-maker's decision rules,  information processing strategies, consideration sets and utility functions. Various approaches are developed to capture heterogeneity in these aspects, such as  \cite{bacsar2004parameterized}, \cite{hess2012allowing}, \cite{vancrane2019}, \cite{barseghyan2021discrete}, and \cite{CRAWFORD20214}. 
Under the most widely used random utility maximization paradigm, how to characterize taste heterogeneity in the utility specification has been a main research topic. The systematic part of a utility function describes how a choice-maker values an alternative  attribute (``taste''), and how tastes vary across choice-makers (``taste heterogeneity"), usually through the interactions between alternative attributes and individual characteristics. Correctly specifying systematic taste heterogeneity can be challenging, especially when the choice problem is novel or complicated without a \textit{priori} knowledge. Misspecified utility functions can lead to lower predictability, biased estimates and misinformed decisions  \citep{bentz2000neural, torres2011wrong, VANDERPOL2014}. 

Nonlinear functions (e.g., higher-order polynomial, semi-log transform, piecewise linear) can characterize systematic taste heterogeneity in more flexible ways. However, it is unwieldy to specify such utility functions as there are many possible nonlinear transformations to be tested. Advanced DCMs, such as the Mixed Logit Model and the Latent Class Choice Model, are developed to model random/unobserved taste heterogeneity not captured by the systematic utility \citep{mcfadden2000mixed, hensher2003mixed, Gupta1994, gopinath1995}. These models substantially enrich taste heterogeneity representation and improve predictability. However, they require both the systematic utility and the random error structure known as a \textit{priori}. Capturing random heterogeneity alone does not resolve the bias induced by misspecified systematic utilities \citep{ben2002hybrid}.

In the machine learning (ML) community, Neural Networks (NN) have become the state-of-art method in many computer vision and natural language processing applications due to superior predictability compared to theory-driven models. Neural networks are shown as a universal function approximator \citep{cybenko1989approximation, hornik1991approximation}, having the ability to learn any arbitrary function from data. This empowers them to learn flexible representation from large datasets, the complexity of which can be difficult to achieve with hand-engineered features by domain experts. 

Numerous studies have examined NN's predictability in the context of mobility choice and shown higher accuracy. However, a significant limitation of NN is the lack of interpretability. Interpretability is crucial for high-stakes decisions in transportation planning, such as infrastructure investment and congestion pricing. The model should represent the real relationships between explanatory variables and choice outcomes and provide reliable answers to ``what-if'' questions at the \textit{disaggregated} level. 

Recent studies utilize NN as a tool to learn flexible behavior representation within the DCM framework while keeping model interpretability \citep{vancrane2019, han2019, SIFRINGER2020236}. The key idea is injecting a neural network into a DCM such that the neural network learns part of the model specification, while the base model is a DCM. We call this type of design the ``Neural-embedded Discrete Choice Model" (NEDCM) \citep{han2019}. \citet{SIFRINGER2020236} develop the first type of such models (L-MNL and L-NL), using neural networks to learn a data-driven part of the utility specification. 

We extend \citet{SIFRINGER2020236}'s L-MNL to a structure called TasteNet-MNL. A neural network (TasteNet) is embedded in an MNL to learn representations of systematic taste heterogeneity. Specifically, a set of taste coefficients in a utility function are learned by a neural network as functions of individual characteristics. Taste coefficients predicted by the neural network are fed into an MNL with explicitly defined utilities. Unknown parameters of the TasteNet and the MNL are jointly estimated. We formalize and strengthen the model interpretability condition. We show that disaggregated interpretability can be achieved by a constrained network architecture and parameter transform that encodes expert knowledge. The source code is made publicly available\footnote{https://github.com/YafeiHan-MIT/TasteNet-MNL}. 

We organize the rest of this paper as follows. Section 2 reviews previous NN applications to travel choice prediction, learning behavior representation, and the challenge of interpretability. Section 3 describes the Taste-MNL structure and implementation. Section 4 shows that TasteNet-MNL can recover the underlying taste heterogeneity and outperform MNLs with under-specified utilities. Section 5 applies TasteNet-MNL to the Swissmetro dataset and shows its higher predictability than MNLs and Mixed Logit benchmarks, and its interpretability at the disaggregated level. Lastly, we discuss our key contributions, limitations of the model, and future works. 

\section{Literature Review}\label{sec:literature}

\subsection{Predictability}

Neural networks have been long treated as a black-box for travel choice prediction. The majority of these studies find the examined NN outperforms a certain type of DCM, such as MNL \citep{Kumar1995, agrawal1996market, west1997comparative, Carvalho1998forecasting, omrani2015predicting, Lee2018, LHERITIER2019198}, nested logit  \citep{Mohammadian2002, cantarella2005, wang2020multitask}, mixed logit \citep{wang2021} and latent class logit model \citep{LHERITIER2019198}. A few studies find similar prediction performance of NN in comparison to nested logit \citep{hensher2000comparison}, MNL \citep{sayed2000comparison, zhao2018MLvsLogit} and mixed logit \citep{zhao2018MLvsLogit}. Neural network's superior predictability is attributed to its universal function approximation capacity \citep{hornik1991approximation}: a NN is able to learn complex nonlinear functions, which can be under-represented in a parametric or semi-parametric DCM setting. 

However, a deep neural network (DNN) with many hidden layers is prone to over-fitting and poor performance on test datasets. Several studies show that increasing the number of hidden layers does not improve predictability and can rather harm it. \cite{nam2017model} shows similar log-likelihood of a DNN compared to the nested logit and cross-nested logit baselines. \cite{wong2019reslogit} find a feed-forward DNN suffers from over-fitting as the number of hidden layers increases. With 4 or more hidden layers, log-likelihood on the validation dataset no longer improves and performs worse than the MNL benchmark. In addition to over-fitting, a DNN often experiences large variations across model runs \citep{glorot2011deep}. In a recent study, \cite{wang2021} conduct a systematic comparison of 11 ML classifiers with 3 types of DCM (logit, nested logit and mixed logit) on 3 datasets with varying sample sizes. They find DNNs and ensemble methods achieve the highest predictability on average and outperform DCMs  by 3-4\%. This study shows that an overly complex DNN or DNNs with poorly tuned hyper-parameters can under-perform a DCM due to its large estimation error. 

To prevent a NN from over-fitting and improve its predictability, regularization strategies such as l2 regularization and drop-out, are commonly employed for NN training. Alternatively, more constrained network architectures are proposed with domain knowledge. \cite{WANG2020234} use alternative-specific networks and find improved predictability compared to fully-connected DNNs. \cite{SIFRINGER2020236}'s L-MNL integrates a NN with a logit model, and uses NN to learn part of the utility. Though interpretability and representation learning is their main motivations, the more customized NN structure can be seen as a regularization strategy. Our study presents a new way to integrate NN and MNL, where NN is utilized to learn taste heterogeneity and constrained by the MNL. We also introduce parameter constraints to incorporate expert knowledge and improve predictability.

\subsection{Interpretability}
A major criticism against NN for travel demand modeling is its black-box nature and a lack of \textbf{interpretability}. 
However, the term ``interpretability" has not been clearly defined or agreed upon.  Interpretability has been a hot topic in the machine learning community, \cite{lipton2017mythos} summarizes different motivations behind the quest for interpretability, such as trust,  understandability, transparency, causality, transferability, informativeness, and fairness. He classifies the properties and techniques that render a model interpretable into two general categories: \textbf{transparency} - to what extent a model can be understood by human, and \textbf{post-hoc interpretability} - the ability to derive useful information from a model. We find this framework helpful to reflect on what kind of ``interpretability" is needed for transportation applications. In the same spirit, \cite{alwosheel2020trustworthy} categorizes the effort to make NN interpretable as: the ante-hoc approach - incorporating interpretability into the model structure; and the post-hoc approach - deriving useful information from models. As interpretability researches spread across many disciplines, and our study is not primarily dedicated to interpretability, our review focuses on interpretability of NNs for travel demand analysis.

The conventional view of NN as a ``black-box" \citep{hensher2000comparison, karlaftis2011statistical} follows the ``transparency" type of interpretability: A NN offers no direct behavioral meaning from the parameters, and it is difficult for a human to understand how it makes certain predictions. In a similar vein, 
\cite{SIFRINGER2020236} considers only the parametric part of the utility as interpretable, and associate interpretability with obtaining unbiased estimates for the transparent part. 

Perhaps a more popular view of interpretability is the ``post-hoc" kind - having the ability to obtain useful information from a model \citep{zhao2018MLvsLogit, WANG2020234}. Researches from early days have shown the analogy between a logit model and an FFW, and the ways to derive economic information from NNs \citep{bentz2000neural, hruschka2002flexible, hruschka2004empirical}. \cite{CHIANG2006514},   \cite{HAGENAUER2017273} and \cite{GOLSHANI201821} conduct sensitivity analysis to measure variable importance. \cite{WANG2020extra} show how to extract a full list of economic indicators from NNs.

Though we have plenty of tools to interpret a NN, a major concern in applying NN for travel demand analysis is the credibility and consistency of its interpretations. \citet{WANG2020extra}  note that economic indicators aggregated over model runs make sense, but results from particular training runs or for individual observations can be unreasonable. For example, choice probability can be non-monotonically decreasing as cost increases and highly sensitive to a particular model run. The derivative of choice probabilities with respect to cost and time can be positive; and value-of-time can be negative, zero, arbitrarily large, or infinite. 

Scenario analysis and policy decision in the field of transportation depends on answers to ``what-if'' questions at the disaggregated level, such as how individuals with specific characteristics would respond to a toll increase. Prediction performance alone does not guarantee correct answers to these questions, as a predictive NN does not necessarily learn the underlying relationships. With an emphasis on trustworthiness, \cite{alwosheel2020trustworthy} frames interpretability as 
to what degree the NN learns intuitive and desirable relations and hence can be trusted. He adapts the method of prototypical examples used in computer vision to diagnose NNs' trustworthiness for discrete choice analysis. 

We share a similar emphasis on trustworthiness in addition to post-hoc information extraction. We define interpretability for transportation applications as follows:

\textit{A model is interpretable if, at the disaggregated level, it can provide credible answers to ``what will happen if'' questions}.
We emphasize the \textbf{credibility} of the economic indicators and interpretability at the \textbf{disaggregated} (both model and choice-maker) level. By ``credible'', we mean the answer should conform with a set of prior knowledge, for example, non-positive choice elasticity regarding cost and non-positive value-of-time. Prior knowledge may change over time and vary across application contexts.

The main interpretability difficulty for neural networks arises from the fact that many NNs can fit the data equally well, but not all can provide realistic behavior interpretations. To improve interpretability, we design a hybrid structure consisting of a neural network and an MNL, and add parameter constraints to reflect expert knowledge. We show that TasteNet-MNL can obtain realistic behavioral indicators at the disaggregated level, and that predictability does not necessarily come at the cost of interpretability.

\subsection{Neural networks for learning behavior representation}
Beyond the prediction focus in most NN applications, an interesting avenue of studies attempts NN to uncover the complex relationships in choice behaviors  \citep{west1997comparative, Carvalho1998forecasting, bentz2000neural, Wong2018, vancrane2019}. For example, \citet{Wong2018} use a restricted Boltzman Machine to capture latent behavior attributes. \cite{vancrane2019} 
develop a NN to learn decision rule heterogeneity. 

There has been a continuing interest in learning flexible utility forms and behavior interpretations with NNs, which is the direct precedent of our study. Early studies conduct Monte-Carlo experiments to show a neural network can capture nonlinearity in utility functions \citep{west1997comparative, Carvalho1998forecasting, bentz2000neural}. Non-linearity may reflect the saturation effect or threshold effect of attributes on utility, or non-compensatory decision rules. For example, \cite{west1997comparative} find that NNs consistently outperform logit and discriminative analysis when predicting the outcome of non-compensatory choice rules. \cite{bentz2000neural} show the analogy between NN and MNL, and NN with hidden layers as a more general version of MNL. With synthetic data and an empirical study, they show that a NN can detect interaction and threshold effects in utility, and therefore can be used as a diagnostic tool to improve MNL utility specification. This sequential approach requires manual analysis of NN results to identify the nonlinear effect, and thus applies only to simple problems.  \cite{hruschka2002flexible} compare a NN with an MNL and a Latent Class Logit (LCL) model in an empirical study of brand choice, and find the NN model can identify interaction effects, threshold effects, saturation effects and other nonlinear forms (like inverse S-shape) of attributes on brand utility. A follow-up study by \cite{hruschka2004empirical} compares NN with two other MNLs with flexible systematic utility, and draws similar conclusions. These studies show that a NN can outperform an MNL, when the nonlinearity in attributes are neglected or mistaken. However, these studies have not addressed nonlinearity in \textit{taste}. They consider NN as either an alternative to MNL; or a diagnostic tool to improve the utility specification of MNL, which works only for simple problems. 

Inspired by the study by \citet{bentz2000neural}, \citet{SIFRINGER2020236} propose to integrate NN and DCM, and learn utility representations with the NN. They divide the systematic utility into a ``knowledge-driven'' part and a ``data-driven'' part learned by a neural network. This structure reduces bias in estimating the knowledge-driven part of the utility, and enhances predictability while maintaining interpretability. Applying this idea to logit and nested-logit, they propose the L-MNL and L-NL structure. 

We extend L-MNL to TasteNet-MNL, using a NN to learn flexible representation of taste functions. In L-MNL, the data-driven part and the knowledge-driven part of the utility are \textit{added}. Variables in the two components neither overlap nor interact with each other. This additive structure can be restrictive since the unused features can interact with variables in the knowledge-driven part of the utility. A common scenario is taste heterogeneity - certain individual characteristics interacting with alternative attributes.  L-MNL can be viewed as learning flexible representations of the Alternative Specific Constants (ASCs) as functions of the unused features (individual characteristics and/or attributes of alternatives), though the data-driven part of the utility it learns is no longer a ``constant" per alternative, but an Individual Specific Constant per alternative. 

TasteNet-MNL differs from L-MNL and traditional FFWs in three aspects. First, we allow all or a subset of taste parameters to be modeled by an FFW as a flexible function. Second, we impose constraints on taste parameters generated by the neural network according to prior knowledge, to regularize the network and obtain interpretable results. Third, we model taste parameters instead of the systematic utilities by a NN, different from the conventional application of an FFW. The key idea is to assign the more complicated or less known task (taste heterogeneity) to a NN while keeping the well-known part parametric.

\section{Model Structure}\label{sec:model}

\textbf{Problem Setup} Suppose for a choice task, each of N individuals makes a one-time choice from an individual-specific choice set $C_n$\footnote{This problem setup can be generalized to repeated choice. We choose one-time choice to simplify the notation}. Observed data for individual $n$ includes personal characteristics $\bm{z}_n$, attributes of each alternative $\bm{x_{in}} (\forall i\in C_n$), and the choice $y_n$. The conditional probability of person $n$ choosing alternative $i$ is $P(y_n=i|\bm{z_n}, \bm{x_{jn}} \forall j \in C_n)$.

If tastes are homogeneous across individuals, the systematic utility ($V_{in}$) of alternative $i$ to a person $n$ can be specified as a linear combination of $K_i$ attributes (Eqn. \ref{eqn:util_mnl_homo}), where $\beta_{ki}$ represents homogeneous \textbf{taste} for attribute $x_{ki}$; and $\beta_{0i}$ is the alternative-specific constant for alternative $i$. 

More realistically, tastes vary across individuals. A systematic utility typically includes interactions between alternative attributes and individual characteristics. Eqn. \ref{eqn:util_mnl_hete} shows such an example, where $\gamma_{pqi}$ represents the interaction effect between the p-th attribute of alternative i ($x_{pin}$), and the q-th characteristics ($z_{qn}$) of individual $n$; and $I_i$ is a set of such pairwise interactions between alternative i's attributes and individual characteristics. Interaction effects are specified based on theory or hypothesis. Given numerous function forms and set of interactions, it can be challenging to test all possible scenarios.

\begin{equation}
\label{eqn:util_mnl_homo}
    V_{in} = \beta_{0i} + \sum_{k=1}^{K_i}{\beta_{ki}x_{kin}}
\end{equation}
\begin{equation}
\label{eqn:util_mnl_hete}
    V_{in} = \beta_{0i} + \sum_{k=1}^{K_i}{\beta_{ki}x_{kin}} + \sum_{(p,q)\in I_i}{\gamma_{pqi}x_{pin}z_{qn}}
\end{equation}

\subsection{Learn systematic taste heterogeneity with neural networks}
We use a neural network to model the interactions between alternative attributes and individual characteristics.  The overall TasetNet-MNL model consists of two modules: a neural network (TasteNet) and a choice module (MNL). 

TasteNet models individual $n$'s taste parameters as flexible functions of this person's characteristics (Eqn. \ref{eqn:beta_tastenet}). The network takes person $n$'s characteristics ($\bm{z_n}$) as inputs, and outputs a vector of taste coefficients $\bm{\beta_n}^{TN}$. $\bm{w}$ is the weights of the network. $\bm{\beta_n}^{TN}$ correspond to all or a subset of taste coefficients. The meaning of each element of $\bm{\beta_n}^{TN}$ is determined by the utility specification in the MNL choice module. The elements of $\bm{\beta_n}^{TN}$ can be alternative-specific, such as mode-specific travel time coefficients in a travel mode choice problem.

\begin{equation}
    \label{eqn:beta_tastenet}
    \bm{\beta}_n^{TN} = TasteNet(\bm{z_n};\bm{w})
\end{equation}

We specify the choice module as a logit model, which can be extended to other model forms such as nested logit. The systematic utility of alternative $i$ is divided into two parts: 1) a flexible part, defined as a dot product of taste coefficients predicted by TasteNet and alternative $i$'s corresponding attributes with flexible coefficients ($\bm{x_{in}^{TN}}$, TN for TasteNet); and 2) a parametric part, a utility function $f$ with attributes $\bm{x_{in}^{MNL}}$ and characteristics $\bm{z_{n}^{MNL}}$ as inputs and $\bm{\beta_{i}^{MNL}}$ as parameters (Eqn. \ref{eqn:util_tastenet}).

According to the condition to obtain unbiased estimates for the MNL part given by \cite{SIFRINGER2020236}, inputs to the neural network should not overlap with variables entering the parametric part of the utility. TasteNet by definition only takes individual characteristics as inputs, and thus separates them from attributes entering the parametric part ($\bm{z_n^{TN}} \cap \bm{x_{in}^{MNL}} = \emptyset$). Additionally, TasteNet and MNL should not share individual characteristics ($\bm{z_n^{TN}} \cap \bm{z_{n}^{MNL}} = \emptyset$). 


The extent to which we let a neural network learn the taste function is up to the modeler. In principle, when we lack confidence about how certain tastes vary among individuals, we let TasteNet learn. Each taste coefficient is \textit{either} predicted by TasteNet \textit{or} estimated in the parametric utility. This is not necessary but desirable for clarity. When a taste function is learned by both TasteNet and MNL, the TasteNet will learn the ``residual" part of the function and should be combined with the parametric part to recover the full taste function. 

\begin{figure}
  \includegraphics[width=\linewidth]{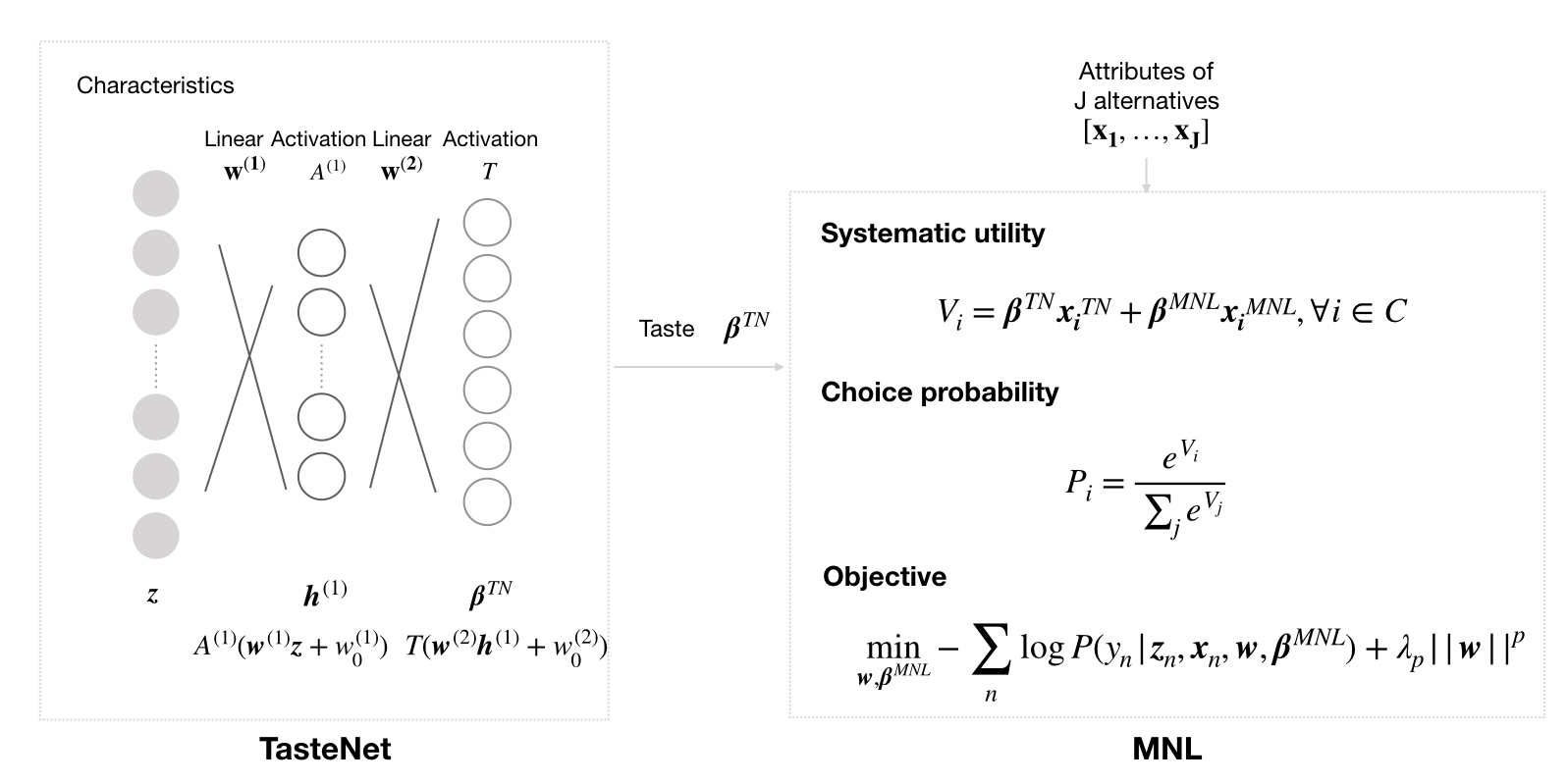}
  \caption{Diagram of a TasteNet-MNL (TasteNet as a feed-forward neural network with 1 hidden layer)}
  \label{fig:tastenet_mnl}
  \centering
\end{figure}

Figure \ref{fig:tastenet_mnl} depicts the overall structure of the TasteNet-MNL. TasteNet learns a function mapping from individual characteristics to tastes. The MNL module combines the flexible and the parametric utility to compute the choice probability for each alternative (Eqn. \ref{eqn:prob}) and choice likelihood. The unknown parameters to estimate are neural network weights ($\bm{w}$) and coefficients ($\bm{\beta}^{MNL}$) in the parametric utilities. They are learned jointly by maximizing the likelihood function.  

\begin{equation}
    \label{eqn:util_tastenet}
    V_{in} = {TasteNet(\bm{z_n}^{TN}; \bm{w})}_{i}^{'}\bm{x_{in}}^{TN} + f(\bm{x_{in}}^{MNL}, \bm{z_n}^{MNL}; \bm{\beta_i}^{MNL})
\end{equation}

\begin{equation}
    \label{eqn:prob}
    P(y_n=i|\bm{x_n},\bm{z_n},\bm{w},\bm{\beta}^{MNL}) = \frac{e^{{TasteNet(\bm{z_n}^{TN};\bm{w})}_{i}^{'}\bm{x_{in}}^{TN} + f(\bm{x_{in}}^{MNL}, \bm{z_n}^{MNL}; \bm{\beta_i}^{MNL})}}
    {\sum_{j\in{C_{n}}}{e^{{TasteNet(\bm{z_n}^{TN};\bm{w})}_{j}^{'}\bm{x_{jn}}^{TN} + f(\bm{x_{jn}}^{MNL}, \bm{z_n}^{MNL}; \bm{\beta_j}^{MNL})}}}
\end{equation}

TasteNet-MNL is distinguished from previous studies in several ways. First, it extends the L-MNL \citep{SIFRINGER2020236} by using a neural network to learn the interactions between characteristics and attributes. Second, different from the majority of neural network applications to discrete choice, TasteNet learns a representation of \textit{taste} rather than utility. This gives us direct control over parameters that carry behavioral meanings, such as value-of-time (VOT), as they are no longer part of the parameters to estimate, but intermediate outputs to \textit{predict}. Thirdly, we realize the necessity of incorporating domain knowledge to obtain interpretable results from neural networks. We introduce parameter constraints along with $l2$ regularization to combat over-parameterization, a common issue with insufficient data and a cause for large estimation variability. Avoiding over-fitting of TasteNet allows for correct learning of both the taste functions and parameters of the utility functions. Hyper-parameter search and model diagnostics should be carefully conducted to avoid over-fitting and ensure stable and interpretable results. 

The rest of this section provides more details on the network architecture, parameter constraints, and estimation. 

\subsection{TasteNet}
\label{sec:tastenet}
The simplest neural network structure is the feed-forward neural network (also called a multi-layer perceptron (MLP)) for TasteNet. An MLP consists of an input layer, one or more hidden layers and an output layer. Essentially, MLP is a composition of linear and nonlinear functions to map inputs to outputs.

In an MLP with 1 hidden layer of H hidden units, the k-th output of the network $\beta_k^{^{TN}}$ can be written as Eqn. \ref{eqn:mlp}, where $D$ is the input dimension, H is the number of hidden units, $A^{(1)}$ is the first hidden layer's activation function, and T is the output layer's activation function. Neural network parameters $\bm{w}^{(1)}$ and $\bm{w}^{(2)}$ correspond to weights of connections from input layer to hidden layer 1, and weights of connections from hidden layer 1 to output layer. 

\begin{equation}
    \label{eqn:mlp}
    \beta_k^{TN} (\bm{z}, \bm{w}) = T[\sum_{h=1}^{H} w_{kh}^{(2)} A^{(1)}(\sum_{i=1}^{D}w_{hi}^{(1)}z_{i}+w_{h0}^{(1)}) + w_{k0}^{(2)}]
\end{equation} 

A general MLP with $L$ hidden layers can be denoted as MLP$(L,[H_1,..,H_L],[A^{(1)}, ...,A^{(L)}],T)$.  We need to specify the number of hidden layers $L$, the size of each hidden layer $H_l$, activation function for each hidden layer $A^{(l)}$, and output transform function $T$. These hyper-parameters are selected based on a model's prediction performance on a hold-out set for model development. 

\subsection{Parameter constraints}

\label{sec:constraint}
Over-parameterization is common for neural networks, especially when the sample size is relatively small compared to the model complexity. Adding constraint is a method to regularize a neural network.  We impose constraints on taste parameters not only to improve model generalization ability; but also to ensure a reasonable range for taste parameters. 

A typical constraint is on the signs of parameters. For example, the coefficient for travel time or waiting time is usually negative. We incorporate sign constraints through an output transform function $T$. For taste parameter $\beta$s with non-negative sign constraints, choices of $T$ can be the rectified linear function ($ReLU(\beta)$) or exponential function ($\exp(\beta)$). For $\beta$s with non-positive signs, choices of $T$ can be the rectified linear unit $-ReLU(-\beta)$ or  $-\exp(-\beta)$. For $\beta$s without constraints, $T$ is the identity function. Such transformations redistribute the parameters to the desirable range through continuous differentiable functions, which resemble the exponential transform for scale or time coefficient commonly used in a DCM utility function. 

An advantage of using a transform function for parameter sign constraints is that the constraints can be strictly kept. Other methods, such as adding a penalty for constraint violation to the learning objective, cannot enforce the constraint on unseen data. 

\subsection{Estimation}
The model is estimated by optimizing an objective function with stochastic gradient descent. The goal is to minimize a loss function, which is the average of negative log-likelihood plus a regularization term for the p-norm of neural network weights (Eqn. \ref{eqn:objective}) to prevent the model from over-fitting. 
\begin{equation}
    \label{eqn:objective}
    \min_{\bm{w}, \bm{\beta}^{MNL}} - \sum_n \log {P(y_n|\bm{z_n}, \bm{x_{in}} \forall i \in C_n; \bm{w}, \bm{\beta}^{MNL})} + \lambda_p ||\bm{w}||^p
\end{equation}

TasteNet-MNL is trained in an integrated fashion through back-propagation using Adam, one of the most successful stochastic gradient descent algorithms for neural network training \citep{kingma2014adam}. Parameters to estimate include neural network weights ($\bm{w}$) and unknown coefficients in the MNL module ($\bm{\beta}^{MNL}$).

\section{Experiments}
We generate synthetic datasets with an underlying logit model, which contains higher-order interactions between characteristics and attributes in its utility functions. We compare TasteNet-MNL with benchmarking MNLs on the synthetic datasets, expecting that the TasteNet-MNL can improve predictability, reduce bias in parameter estimates, and provide more accurate behavioral interpretations, compared to under-specified MNLs. 

\subsection{Synthetic datasets}

The data generation model is a binary logit, with the systematic utility of alternative $i$ for person $n$ defined in Eqn. \ref{eqn:Vin}. Explanatory variables include three  characteristics: income  ($inc$),  full-time  employment  dummy  ($full$)  and flexible work schedule dummy ($flex$); and three alternative attributes: travel cost ($cost$) and travel time ($time$) and waiting time ($wait$) (see Table \ref{tab:taste_toy_desc} in Appendix \ref{app-A} for details). Values of the coefficients are chosen to be realistic: income has a positive effect on value-of-time (VOT) and value-of-waiting-time (VOWT), full-time workers have higher VOT and VOWT, and people with flexible schedules have lower VOT and VOWT. We intentionally include second-order interactions between individual characteristics and alternative attributes to test whether TasteNet can learn them. We assign -1 as the ground-truth cost coefficients for both alternatives, so that outputs of the TasteNet are in the willingness-to-pay space, corresponding to VOT and VOWT (Eqn. \ref{eqn:Vin}). Alternative specific constant (ASC) for alternative 0 is fixed to 0. The random component of each utility follows an Extreme Value distribution. 

We create two synthetic datasets: TOY\_UNCORREL with uncorrelated alternative attributes; and TOY\_CORREL with a 0.6 correlation between $time$ and $wait$. Having the correlated dataset helps examine whether the neural network can learn multiple taste functions accurately and consistently when the corresponding attributes are correlated.

Each dataset contains 14,000 observations, randomly split into training (10,000), development (2000) and test (2000) sets. Details about the input data distribution and synthetic data generation are reported in Appendix \ref{app-A}.

\begin{multline}
    V_{in}= ASC_i - cost_{in} + \\
    (-0.1 - 0.5inc_n - 0.1full_n + 0.05flex_n \\
    - 0.2inc_n * full_n + 0.05inc_n * flex_n 
    + 0.1full_n * flex_n) * time_{in} + \\
    (-0.2 - 0.8inc_n - 0.3full_n + 0.1flex_n \\
    - 0.3inc_n * full_n + 0.08inc_n * flex_n 
    + 0.3full_n * flex_n) * wait_{in}
\label{eqn:Vin}
\end{multline}

\subsection{Models in comparison}
\label{sec:tastenet_benchmark}

We specify three MNL benchmarks with increasing complexity in the utility specifications. \textbf{MNL-I}'s utility functions only include first-order interactions between individual characteristics and attributes ($time$ and $wait$) (Eqn. \ref{eqn:MNL-I-V}). Compared to MNL-I, utilities of \textbf{MNL-II} have one additional interaction $inc*full*time$ (Eqn. \ref{eqn:MNL-II-V}). \textbf{MNL-TRUE} has the same utility function as the data generation model.

\begin{multline}
V_i^{MNL-I} = ASC_i - cost_i
+ (b_0 + b_1 inc + b_2 full + b_3 flex)*time_i\\
+ (c_0 + c_1 inc + c_2 full + c_3 flex)*wait_i
\label{eqn:MNL-I-V}
\end{multline}

\begin{multline}
V_i^{MNL-II} = ASC_i - cost_i + (b_0 + b_1 inc + b_2 full + b_3 flex + b_{12}inc*full)*time_i \\
+ (c_0 + c_1 inc + c_2 full + c_3 flex + c_{12}inc*full)*wait_i
\label{eqn:MNL-II-V}
\end{multline}

\begin{figure}[!htbp]
  \includegraphics[width=\linewidth]{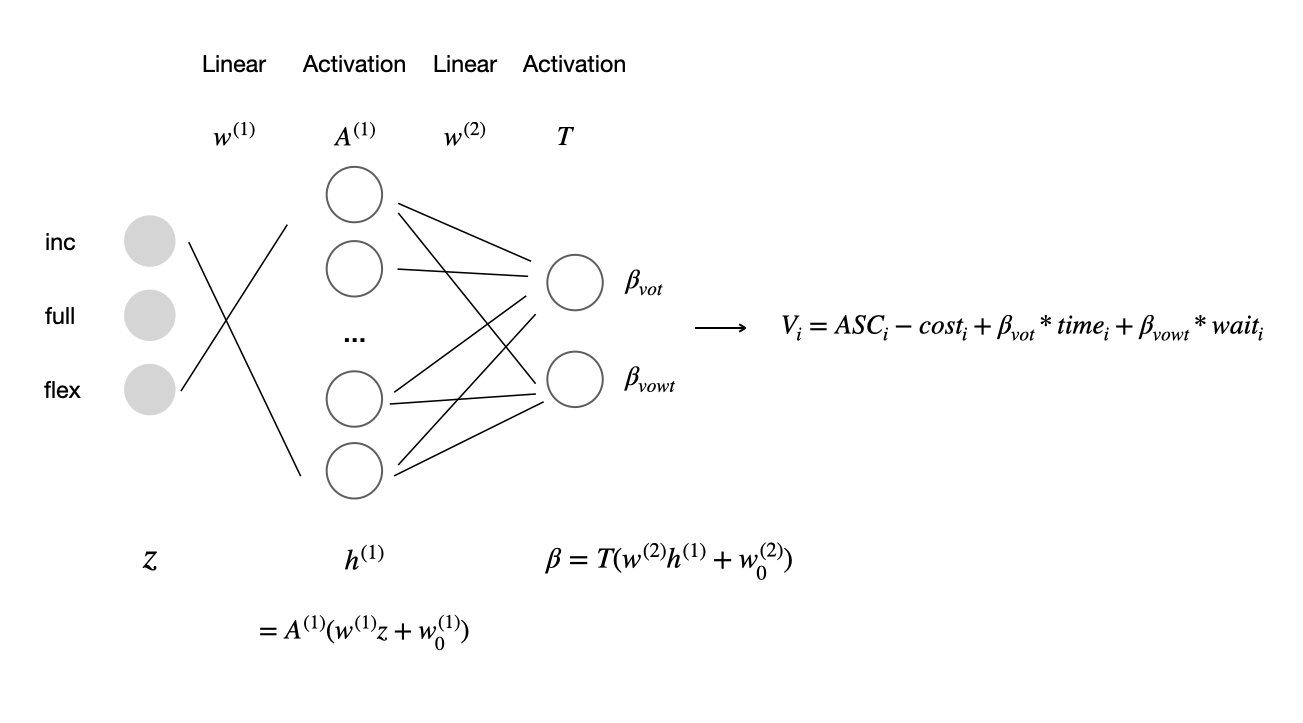}
  \caption{TasteNet-MNL Diagram for the  Synthetic Data}
  \label{fig:taste_toy_diag}
  \centering
\end{figure}

Figure \ref{fig:taste_toy_diag} shows the structure of the TasteNet-MNL for the synthetic data. Time coefficient ($\beta_{time}$) and waiting time coefficient ($\beta_{wait}$ is modeled by an MLP. Hyper-parameters include the number of hidden layers, the size(s) of hidden layers, type of regularizer, regularization strength, activation function for hidden layers, and output transform function. 

We train a TasteNet-MNL model on the training dataset with different combinations of hyper-parameters. One hidden layer turns out to be sufficient to achieve the same accuracy as the data generation model. We vary the number of hidden units from 10 to 100. For each hidden layer size, we apply $l2$ penalty in [0, 0.0001, 0.001, 0.01]. For the hidden layer activation function, we experiment with the ReLU and Tanh function. For output transformation, we experiment with functions: $-ReLU(-\beta)$ and $-e^{-\beta}$, to impose non-positive constraint on the value of time coefficient $\beta_{time}$ and $\beta_{wait}$. For each combination of hyper-parameters, we train the model 100 times with different random initialization. 

\subsection{Results}
The optimal set of hyper-parameters is selected based on the lowest average negative log-likelihood ($NLL$) on the development dataset. On the TOY\_UNCORREL dataset, the best TasteNet-MNL has 1 hidden layer with 60 hidden units, $ReLU$ for hidden layer activation, $-ReLU(-\beta)$ for output transformation, and $l2$ penalty 0.001. On TOY\_CORREL dataset, the optimal TasteNet-MNL has 1 hidden layer with 100 hidden units, $ReLU$ for hidden layer activation, $-ReLU(-\beta)$ for output transformation, and $l2$ penalty 0.001. We choose the smallest layer-depth model with similar performances to larger models to reduce the chance of model overfitting to spurious examples. We compare MNLs and TasteNet-MNL with respect to their predictability, parameter bias, and interpretability. 

\subsubsection{Predictability}

Model predictability is measured by average negative log-likelihood (NLL) and prediction accuracy (ACC) on training, development and test data. NLL is the total negative log-likelihood in equation (\ref{eqn:objective}) divided by the number of observations. The higher the NLL, the poorer the model fit and predicative power. Prediction accuracy is the percentage of correct predictions. Table \ref{tab:taste_toy_nll_acc} and Table \ref{tab:taste_toy_nll_acc_correl} summarize the prediction performance of different models on the synthetic datasets.

\begin{table}[!htbp]
 \caption{Average Negative Log-likelihood (NLL) and Prediction Accuracy (ACC) for Synthetic Data with Uncorrelated Alternative Attributes}
  \centering
  \resizebox{\textwidth}{!}
  {\begin{tabular}{lrrrrrr}
    \toprule
    Model & NLL\_train & NLL\_dev & NLL\_test & ACC\_train & ACC\_dev & ACC\_test \\
    \midrule
    MNL-I & 0.59775 & 0.57773 & 0.57866 &  0.712  & 0.725 & 0.739\\
    MNL-II & 0.50618 & 0.47657 & 0.49603  & 0.755 & 0.768 & 0.760 \\
    TasteNet-MNL \\
    \multicolumn{1}{r}{best run} & 0.46523	& 0.44193 & 0.44293	& 0.777	& 0.785 & 0.789 \\
    \multicolumn{1}{r}{mean}  & 0.46688 &	0.44494	& 0.44536	& 0.775	& 0.787	& 0.785 \\
    \multicolumn{1}{r}{std} & (0.00159) & (0.00141) & (0.00225) &	(0.002) &	(0.002) &	(0.003) \\
     \midrule
     MNL-TRUE  & 0.46424 & 0.44028 & 0.44669 & 0.777 & 0.786 & 0.784 \\
     Ground-truth & 0.46460  & 0.44120 & 0.44655  & 0.775  &  0.786  &  0.788 \\
    \bottomrule
  \end{tabular}}
  \label{tab:taste_toy_nll_acc}
\end{table}
\begin{table}[!htbp]
 \caption{Average Negative Log-likelihood (NLL) and Prediction Accuracy (ACC) for Synthetic Data with Correlated Alternative Attributes}
  \centering
  \resizebox{\textwidth}{!}
  {\begin{tabular}{lrrrrrr}
    \toprule
    Model & NLL\_train & NLL\_dev & NLL\_test & ACC\_train & ACC\_dev & ACC\_test \\
    \midrule
    MNL-I & 0.67756 & 0.62325	& 0.66171 & 0.698 & 0.706 & 0.705 \\
    MNL-II & 0.53260 & 0.52495	& 0.53488 & 0.748 & 0.741 & 0.750 \\
    TasteNet-MNL \\
    \multicolumn{1}{r}{best run} & 0.46261	& 0.46671	& 0.45345	& 0.777	& 0.773	& 0.776\\
    \multicolumn{1}{r}{mean}  & 0.46125	& 0.47171 & 0.45332	& 0.777	& 0.768	& 0.776 \\
    \multicolumn{1}{r}{std} & (0.00198) & (0.00216)	& (0.00275) & (0.002)	& (0.003) & (0.003) \\
    \midrule
    MNL-TRUE  & 0.45839	& 0.47018 & 0.45228	& 0.779	& 0.768 & 0.778\\
    Ground-truth & 0.45945 & 0.46845 & 0.45110 & 0.777 & 0.767 & 0.781 \\
    \bottomrule
  \end{tabular}}
  \label{tab:taste_toy_nll_acc_correl}
\end{table}

On TOY\_UNCORREL, MNL-I and MNL-II result in higher NLL (0.58, 0.5) than MNL-TRUE (0.45); and lower prediction accuracy (74\%, 76\%) than MNL-TRUE (78\%) due to missing interaction terms. 
The performance gap is widened on the TOY\_CORREL dataset: MNL-I and MNL-II achieve higher NLL (0.66, 0.53) compared to MNL-TRUE (0.45) and lower accuracy (71\%, 75\%) compared to the true model (78\%). On both datasets, TasteNet-MNL's  best estimation run, as well as the average of the 100 estimation runs reach the same level of predictability as MNL-TRUE. The small standard deviation suggests its prediction performance is stable and significantly higher than the under-specified MNLs.


Readers may wonder: does TasteNet-MNL discover the underlying utility function form? Is this predictability gain attributed to TasteNet's ability to \textit{learn the underlying taste functions} or simply a result of over-fitting? The next section examines the taste functions learned by TasteNet and compare it with the underlying model. 

\subsubsection{Parameter Estimates}
Coefficients in the MNL utility functions describe the marginal effect of a unit change in the inputs. 
A neural network does not provide direct estimates of the coefficients. In this experiment, because we know the terms in the underlying utility functions, we can regress predicted $\beta_{time}$ and $\beta_{wait}$ for all individuals by TasteNet against these terms to compare the coefficient estimates with the ground-truth. This measures how accurately the TasteNet recovers the underlying taste functions. The alternative specific constant ($ASC_1$) is obtained from the MNL component as it is not included in the TasteNet part of the utility. Table \ref{tab:taste_toy_param} and \ref{tab:taste_toy_param_correl} show the coefficient estimates and error metrics by different models for the two toy datasets. 

\begin{table}[!htbp]
 \caption{Parameter Estimates by MNLs and TasteNet-MNL (Uncorrelated Data)}
  \centering
  \begin{threeparttable}
  \resizebox{\textwidth}{!}
  {\begin{tabular}{lrrrrrrr}
    \toprule
    Coef & MNL-I & MNL-II & MNL-TRUE & TasteNet-MNL & TasteNet-MNL & Truth\\
    & &  & & (best run) & (100 runs: mean, std) & \\
    \midrule
    ASC1 & 
    -0.1358 & -0.1263 	& -0.1325 & -0.1315 &  -0.1321	& -0.10 \\
    & (0.0273) & (0.0262) & (0.0257) &
    & (0.0207)\\
    
    b\_time & 
    -0.0875	& -0.0330 & -0.0964	& 	-0.1028	& -0.0983	& -0.10 \\
    & (0.0023) & (0.0027) & (0.0051) & & (0.0090) \\
    
    b\_time\_flex	& 0.1205	& 0.0050	& 0.0501	& 0.0534	& 0.0534	& 0.05 \\
    & (0.0015)	& (0.0035)	& (0.0039)	& & (0.0077)\\
    
    b\_time\_full	 & -0.1070	 & -0.1062	 & -0.1001	 & -0.0969	 & -0.1174	 & -0.10 \\
    & (0.0025)	& (0.0024)	& (0.0065)	& & (0.0156) \\
    
    b\_time\_full\_flex & & & 0.0979	& 0.0823	& 0.0886	& 0.10 \\
    & & & (0.0044)	& & (0.0156) \\

    b\_time\_inc	 & -0.6582	 & -0.8026	 & -0.5176	 & -0.5122	 & -0.5307	 & -0.50  \\
    & (0.0077)	& (0.0083)	& (0.0194)	& & (0.0445) \\
    
    b\_time\_inc\_flex	& & 0.3014	& 0.0555	& 0.0819	& 0.0631	& 0.05 \\
   	& & (0.0085)	& (0.0135)	& & (0.0317) \\
    
    b\_time\_inc\_full & & &  -0.1897	 & -0.1914	 & -0.1447	 & -0.20 \\
    & & & (0.0196)	& & (0.0460) \\
    
    b\_wait	 & -0.2188	 & -0.0625	 & -0.1874	 & -0.1902	 & -0.2091	 & -0.20 \\
    & (0.0085)	& (0.0103)	& (0.0198)	& & (0.0187) \\
    
    b\_wait\_flex	& 0.2834	 & -0.0304	& 0.1042	& 0.1013	& 0.1091	& 0.10 \\
    & (0.0055)	& (0.0133)	& (0.0148)	& & (0.0148) \\

    b\_wait\_full	 & -0.2337	 & -0.2281	 & -0.2976	 & -0.3284	 & -0.3187	 & -0.30 \\
    & (0.0090)	& (0.0090)	& (0.0250)	& & (0.0295) \\
    
    b\_wait\_full\_flex	& &	& 0.2799 & 0.2842 & 0.2454	& 0.30 \\
    & & & (0.0167)	& & (0.0283) \\
    
    b\_wait\_inc	& -1.0335	& -1.4497	 & -0.8641	 & -0.8431	 & -0.8556	 & -0.80 \\
    & (0.0277)	& (0.0318)	& (0.0756)	& & (0.0727) \\
    
    b\_wait\_inc\_flex	& & 0.8148	& 0.0976 & 0.0720	& 0.1278 & 0.08 \\
    & & (0.0319)	& (0.0511)	& & (0.0609) \\
    
    b\_wait\_inc\_full	& & & -0.2557 & -0.2123	 & -0.2041	 & -0.30 \\
    & & & (0.0761)	& & (0.0762) \\
    \midrule
    RMSE & 0.0911   & 0.2787   & 0.0238   &	0.0298	   & 0.0520  \\
    &&&&&(0.0203)\\
    MAE & 0.0524    & 0.1615	   & 0.0158	   & 0.0204	   & 0.0379 \\
    &&&&& (0.0141)\\
    MAPE & 0.3146	   & 1.3172	   & 0.0805	   & 0.1320	   & 0.2359\\
    &&&&& (0.0849)\\
    \bottomrule
  \end{tabular}}
      \begin{tablenotes}[]
         \footnotesize
         \item RMSE: Root Mean Squared Error; MAE: Mean Absolute Error; MAPE: Mean Absolute Percentage Error
    \end{tablenotes}
\end{threeparttable}
  \label{tab:taste_toy_param}
\end{table}
\begin{table}[!htbp]
 \caption{Parameter Estimates by MNLs and TasteNet-MNL (Correlated Data)}
  \centering
  \begin{threeparttable}
  \resizebox{\textwidth}{!}
  {\begin{tabular}{lrrrrrrr}
    \toprule
    Coef & MNL-I & MNL-II & MNL-TRUE & TasteNet-MNL & TasteNet-MNL & Truth\\
    & &  & & best run & 100 runs: mean (std) & \\
    \midrule
    ASC1	&	-0.0335	&	-0.0385	&	-0.0543	&	-0.0561	&	-0.0627	&	-0.10	\\
    &	(0.0282)	&	(0.0265)	&	(0.0258)	&	& (0.0528)	\\
    
    b\_time	&	-0.0871	&	-0.0310	&	-0.0988	&	-0.1090	&	-0.0963	&	-0.10	\\
    &	(0.0028)	&	(0.0033)	&	(0.0063)	&	& (0.0089)	\\
    
    b\_time\_flex	&	0.1211	&	0.0016	&	0.0500	&	0.0723	&	0.0592	&	0.05	\\
    &	(0.0018)	&	(0.0043)	&	(0.0047)	&	&  (0.0088)	\\
    
    b\_time\_full	&	-0.1103	&	-0.1113	&	-0.1003	&	-0.1030	&	-0.1271	&	-0.10	\\
    &	(0.0030)	&	(0.0030)	&	(0.0080)	&	& (0.0174)	\\
    
    b\_time\_full\_flex			&	&	&	0.1012	&	0.0861	&	0.0872	&	0.10	\\
    &		&		&	(0.0054)	&	& (0.0142)	\\
    
    b\_time\_inc	&	-0.6562	&	-0.8004	&	-0.4979	&	-0.4836	&	-0.5439	&	-0.50	\\
    &	(0.0093)	&	(0.0103)	&	(0.0241)	&	& (0.0363)	\\
    
    b\_time\_inc\_flex		&	&	0.3087	&	0.0497	&	0.0248	&	0.0562	&	0.05	\\
    &		&	(0.0104)	&	(0.0161)	&	& (0.0283)	\\
    
    b\_time\_inc\_full			&	&	&	-0.2077	&	-0.1866	&	-0.1279	&	-0.20	\\
    &		&		&	(0.0243)	&	& (0.0410)	\\

    b\_wait	&	-0.2256	&	-0.0663	&	-0.1790	&	-0.2196	&	-0.2079	&	-0.20	\\
    &	(0.0106)	&	(0.0125)	&	(0.0242)	&  & 	(0.0170)	\\

    b\_wait\_flex	&	0.2805	&	-0.0587	&	0.0951	&	0.0834	&	0.1004	&	0.10	\\
    &	(0.0068)	&	(0.0161)	&	(0.0178)	& & 	(0.0173)	\\

    b\_wait\_full	&	-0.1913	&	-0.2068	&	-0.3256	&	-0.3054	&	-0.3243	&	-0.30	\\
    &	(0.0111)	&	(0.0110)	&	(0.0304)	& & 	(0.0283)	\\

    b\_wait\_full\_flex			&	&	&	0.3196	&	0.2856	&	0.2735	&	0.30	\\
    &		&		&	(0.0201)	& & 	(0.0230)	\\

    b\_wait\_inc	&	-1.1014	&	-1.4841	&	-0.9142	&	-0.8038	&	-0.8586	&	-0.80	\\
    &	(0.0340)	&	(0.0385)	&	(0.0922)	& & 	(0.0680)	\\

    b\_wait\_inc\_flex	&	&	0.8868	&	0.0576	&	0.0886	&	0.0974	&	0.08	\\
    &		&	(0.0388)	&	(0.0612)	& & (0.0544)	\\
    
    b\_wait\_inc\_full	&	&	&	-0.1742	&	-0.2500	&	-0.1785	&	-0.30	\\
    &		&		&	(0.0929)	& & 	(0.0654)	\\
    
    \midrule
    RMSE	&	0.1065	&	0.2988	&	0.0469	&	0.0220	&	0.0545\\
	&		&		&		&		& (0.0197) \\
    MAE	&	0.0622	&	0.1751	&	0.0261	&	0.0177	&	0.0391	\\
	&		&		&		&
	& (0.0130) \\
    MAPE	&	0.3537	&	1.4445	&	0.1121	&	0.1571	&	0.2428	\\
    &		&		&		&
	& (0.0706) \\
    \bottomrule
  \end{tabular}}
  
  \begin{tablenotes}[]
             \footnotesize
             \item RMSE: Root Mean Squared Error; MAE: Mean Absolute Error; MAPE: Mean Absolute Percentage Error
        \end{tablenotes}

\end{threeparttable}
  \label{tab:taste_toy_param_correl}
\end{table}

On the TOY\_UNCORREL dataset, TasteNet-MNL's parameter bias is slightly higher than MNL-TRUE, and much lower than the misspecified MNLs. The best run of TasteNet-MNL achieves slightly higher parameter errors (RMSE=0.03, MAE=0.02, MAPE=13.2\%) than MNL-TRUE (RMSE=0.024, MAE=0.016, MAPE=8.1\%). The average bias from 100 TasteNet-MNL runs is higher (RMSE=0.052, MAE=0.038, MAPE=23.6\%) than the best run, but still lower than MNL-I (RMSE=0.09, MAE=0.05, MAPE=31.5\%) and MNL-II (RMSE=0.28, MAE=0.16, MAPE=132\%).

Though MNL-II's utility specification is more complete and closer to the ground-truth, its parameter bias is larger than the simple MNL-I. Adding the second-order interactions (time*inc*flex and wait*inc*flex) causes severe bias in related coefficients, such as b\_time, b\_time\_flex, b\_time\_inc, b\_wait, b\_wait\_flex and b\_wait\_inc. This bigger bias may not be detected by modelers as MNL-II's predictability is higher than MNL-I (Table \ref{tab:taste_toy_nll_acc}). This suggests that unless the exact utility form is correctly specified, there is a risk of having large bias with good predictability. A neural network is able to learn taste functions from data and mitigate this risk. 

With a 0.6 correlation between $time$ and $wait$, all MNLs' parameter bias increase (Table \ref{tab:taste_toy_param_correl}). Standard errors of the coefficients also increase. MNL-TRUE's parameter RMSE increases from 0.0238 to 0.0469, almost doubled. This is because when alternative attributes are highly correlated, the decision boundary becomes less certain. 

With the correlated attributes (time and wait), readers may suspect TasteNet-MNL to suffer from less accurate and more unstable parameter estimates, because the two tastes are estimated by a \textit{shared} neural network, which may interfere with each other. The best run of TasteNet-MNL's parameter bias stays almost at the same level, and the average bias increases by about 3 to 5\%. There is no strong evidence of deteriorated accuracy or stability caused by interactions between the two taste functions. It seems that a linear model is more sensitive to the correlation in the input data. 

Why is it? Our intuition is as follows. Although the two tastes are estimated through a shared neural network, they do not necessarily interact much with each other because 1) each taste output has its own set of weights for the connections between the hidden layer and the output node; and 2) the number of hidden nodes is not too small (restrictive). For example, when the hidden layer is fairly large, taste output 1 might be mainly connected with a subset of the hidden nodes, while taste output 2 can be mainly connected with another subset. With a wide hidden layer, there is enough flexibility to learn two taste parameters that do not interfere with each other and together optimize the objective function.The $l2$ regularization penalty, if properly tuned, can keep the most useful connections, while shrinking the rest towards zero. The weights of the network may vary much across runs (the well-known unidentifiable nature of a neural network), but the output of the network - the taste parameters are constrained by the downstream MNL utility, and therefore can remain relatively stable. Because MNLs are constrained by the overarching linear structure of the utility, there is less freedom to adapt to the changes in the shape of the input distribution. However, with the flexibility of the neural network also comes a weakness: a neural network does not perform well on out-of-distribution samples or samples with low occurrences in the training data. We will discuss this issue in the next section. 

\subsubsection{Taste Estimates and Taste Functions}

We estimate VOT and VOWT for each individual in the synthetic datasets, and compute the errors compared to the ground-truth (Table \ref{tab:taste_toy_vot}). 

On the TOY\_UNCORREL dataset, TasteNet-MNL obtains smaller errors in its predicted VOT and VOWT than MNL-I and MNL-II. TasteNet-MNL's mean absolute error (MAE) for VOT is 0.15\$/hr (0.8\% of the true value), compared to MNL-I's 1.7\$/hr (11\%) and MNL-II's 0.8\$/hr (4.8\%).  TasteNet-MNL's MAE for VOWT is 0.6\$/hr (2\%) compared to MNL-I's 4.8\$/hr (17.7\%) and MNL-II's 2.3\$/hr (7.7\%). 

On the TOY\_CORREL dataset, both MNL-TRUE and TasteNet-MNL's VOT and VOWT's errors increase, but the error magnitudes relative to each other remain the same: TasteNet-MNL's taste error is about 3 times as big as MNL-TRUE's. TasteNet-MNL maintains its significant lead over MNL-I and MNL-II. 

\begin{table}[!htbp]
    \centering
    \caption{Errors of the Estimated Value of Time and Value of Waiting Time (Unit: \$ / Hour)}
    \begin{threeparttable}
    \resizebox{\textwidth}{!}
        {\begin{tabular}{llrrrrr}
        \toprule
        Input data & Error metric\tnote{a} & MNL-I & MNL-II & MNL-TRUE	& TasteNet-MNL & TasteNet-MNL\\
        & &  & & & best run & 100 runs: mean (std)\\
        \midrule
        Uncorrelated Data 
        & 	RMSE\_VOT	& 	1.76	& 	1.00	& 	0.06	& 	0.19	& 	0.25 (0.06)	\\
        & 	MAE\_VOT	& 	1.70	& 	0.80	& 	0.04	& 	0.15	& 	0.20	(0.06)	\\
        & 	MAPE\_VOT	& 	11.0\%	& 	4.8\%	& 	0.3\%	& 	0.8\%	& 	1.2\% (0.4\%)	\\
        & 	RMSE\_VOWT	& 	4.87	& 	2.81	& 	0.28	& 	0.72	& 	0.79 (0.28)	\\
        & 	MAE\_VOWT	& 	4.82	& 	2.31	& 	0.24	& 	0.60	& 	0.66 (0.29)	\\
        & 	MAPE\_VOWT	& 	17.7\%	& 	7.7\%	& 	0.9\%	& 	2.0\%	& 	2.3\% (1.0\%) \\
        \midrule
        Correlated Data
        & 	RMSE\_VOT	& 	1.76	& 	1.01	& 	0.10	& 	0.36	& 	0.29	(0.08)	\\
        & 	MAE\_VOT	& 	1.70	& 	0.81	& 	0.09	& 	0.30	& 	0.23	(0.08)	\\
        & 	MAPE\_VOT	& 	11.2\%	& 	4.9\%	& 	0.6\%	& 	2.1\%	& 	1.4\%	(0.5\%)	\\
        & 	RMSE\_VOWT	& 	4.97	& 	2.94	& 	0.68	& 	1.27	& 	1.02	(0.29)	\\
        & 	MAE\_VOWT	& 	4.82	& 	2.17	& 	0.49	& 	0.97	& 	0.86	(0.26)	\\
        & 	MAPE\_VOWT	& 	16.9\%	& 	6.3\%	& 	2.3\%	& 	4.8\%	& 	3.4\%	(1.2\%)	\\
    	\bottomrule
        \end{tabular}}
        \begin{tablenotes}[]
         \footnotesize
         \item RMSE: Root Mean Squared Error; MAE: Mean Absolute Error; MAPE: Mean Absolute Percentage Error\\
        VOT: Value of Time;
        VOWT: Value of Waiting Time
        \end{tablenotes}
    \end{threeparttable}
    \label{tab:taste_toy_vot}
\end{table}

\begin{figure}[!htbp]
  \includegraphics[width=\textwidth]{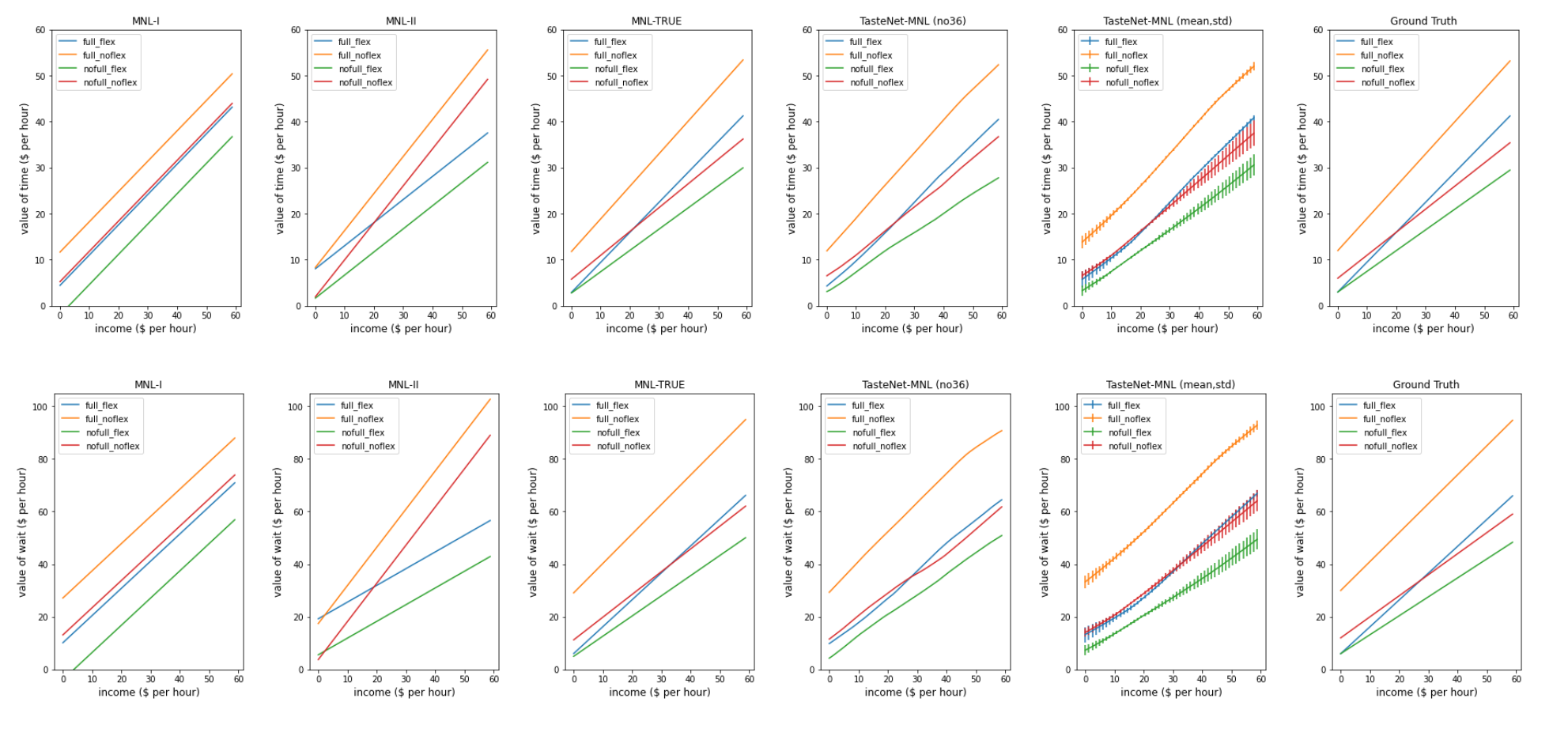}
  \centering
  \caption{Estimated Value of Time and Value of Waiting Time v.s. Income for 4 Groups (data: TOY\_UNCORREL)}
  \label{fig:taste_uncorrel}
  \centering
\end{figure}

\begin{figure}[!htbp]
  \includegraphics[width=\textwidth]{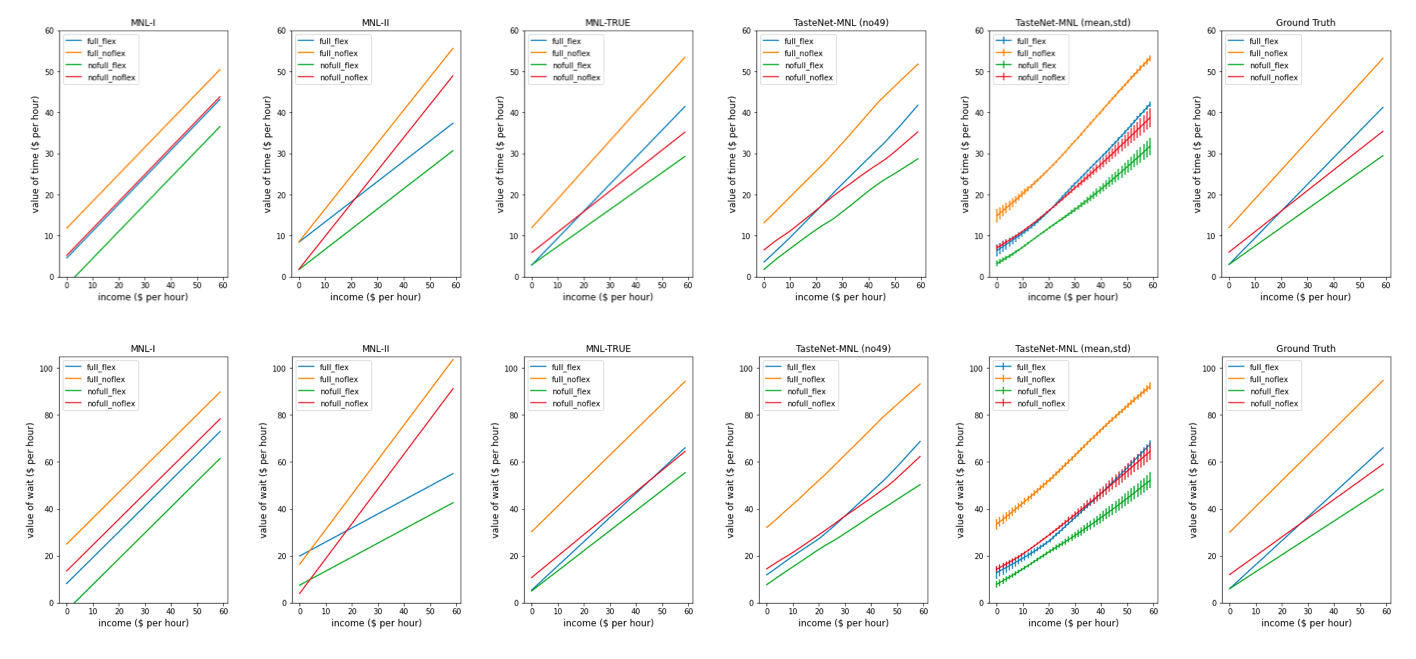}
  \centering
  \caption{Estimated Value of Time and Value of Waiting Time v.s. Income for 4 Groups (data: TOY\_CORREL)}
  \label{fig:taste_correl}
  \centering
\end{figure}

To further investigate the shape of the taste functions learned by the neural network compared with other models, we create a dataset with 200 individuals with uniformly distributed characteristics. Note that this input distribution is different from that of the training data. 50 individuals are in each of the four groups defined by all combinations of full-time (yes/no) and flexible schedule (yes/no). The income of individuals from each group is evenly distributed from 0 to 60\$ per hour with a step size 1.2. We plot the predicted VOTs and VOWTs against the continuous income for the four types of individuals. Figure \ref{fig:taste_uncorrel} shows the results from applying models for the TOY\_UNCORREL dataset and Figure \ref{fig:taste_correl} shows the results from applying models for the TOY\_CORREL dataset. The ground-truth utility functions are the same between the two. 

MNL-I cannot distinguish the differences in the shape of VOT and VOWT functions between the \textit{full\_flex} group and the \textit{nofull\_noflex} group, due to missing all 3 second-order interaction terms. Adding two second-order interactions to MNL-II causes larger bias, and significant changes in the slope of the VOT and VOWT functions of income for each group. 

Unsurprisingly MNL-TRUE's estimated taste functions are the closest to the ground-truth, as its utility has the exact terms as the underlying utility. Comparing MNL-TRUE's taste functions between uncorrelated data (Figure \ref{fig:taste_uncorrel}) and correlated data (Figure \ref{fig:taste_correl}), we notice the  MNL-TRUE's VOWT function has an apparent deviation from the ground-truth due to the correlation in attributes. 

\begin{figure}[!htbp]
  \includegraphics[width=0.6\textwidth]{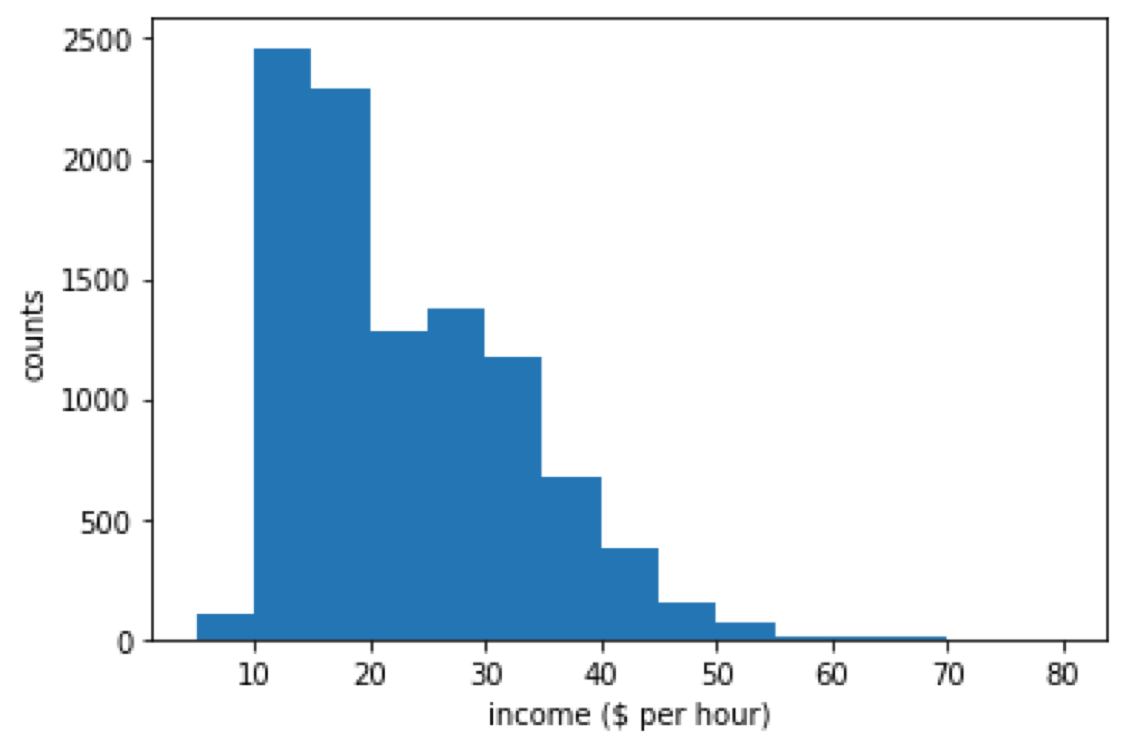}
  \centering
  \caption{Histogram of Income in Training Data}
  \label{fig:income_hist}
  \centering
\end{figure}

TasteNet-MNL overall recovers the true shape of the taste functions for both the uncorrelated and the correlated case. However, the functions are not strictly linear. Near the limits of income, we observe larger bias and variance. This is because the majority of the training samples have income between 10 and 50 \$/hr (Figure \ref{fig:income_hist}), with only 208 (2.1\%) of the 10,000 training samples beyond this range. With insufficient examples in a certain region of the input space, the neural network is unable to learn an accurate and precise taste function. This reveals a major weakness of TasteNet-MNL that it does not generalize well to out-of-distribution samples. 

Despite this weakness, TasteNet-MNL has the ability to recover the underlying taste functions for the majority of the population, given that there is enough data to learn from. It can lower the risk of getting the taste functions systematically wrong without knowing the true utility form. Whether the data is sufficient depends on the complexity of the problem. An MNL model tends to generalize better to out-of-distribution samples because of the highly restricted global utility functions. However, this generalization ability is only possible if we have a correct specification of the utility functions. As the MNL-II example shows, a misspecified MNL can lead to greater bias for the out-of-distribution samples than a TasteNet-MNL does. 

It is worth noting that in this synthetic data experiment, we examine a particular form of nonlinearity - higher-order interactions - for the convenience of illustration and understanding. Based on the universal function approximation property of NN, TasteNet can be employed to learn general forms of nonlinearity in taste functions. 

\section{Model Application: Swissmetro Mode Choice}
We apply TasteNet-MNL to a publicly available dataset -- \textit{Swissmetro} to model mode choice for inter-city travel. The purpose of this application is to 1) examine whether TasteNet-MNL is able to predict more accurately compared to a manually specified, relatively sophisticated MNL; and 2) whether TasteNet-MNL can draw reasonable behavioral interpretations and, if so, how its interpretations differ from those of the MNLs. To compare with TasteNet-MNL, we set up three benchmarking MNL models with increasing complexity in the utility function and random coefficient logit 
models. 

\subsection{Data}
\label{sec:sm_data}
The Swissmetro is a proposed revolutionary mag-lev underground system. To assess potential demand, the Swissmetro Stated Preference (SP) survey collected data from 1,192 respondents (441 rail-based travelers and 751 car users), with 9 choices from each respondent. Each respondent is asked to choose one mode out of a set of alternatives for inter-city travel given the attributes of each mode (e.g. travel time, headway and cost). The universal choice set includes train (TRAIN), Swissmetro (SM), and car (CAR). For individuals without a car, the choice set includes only TRAIN and SM. Table \ref{tab:sm_data} provides a description of the variables. For more information, readers can refer to \cite{bierlaire2018}.

The original data has 10,728 observations, downloaded in Jan 2019\footnote{Data link: https://biogeme.epfl.ch/data.html}. After removing observations with unknown age, ``other" trip purpose and unknown choice, we retain 10,692 observations. We randomly split the data into training (``train"), development(``dev") and test(``test") set with 7,484, 1,604 and 1,604 observations, respectively. 

\begin{table}[!htbp]
 \caption{Description of Variables in the Swissmetro Dataset}
 \label{tab:sm_data}
  \centering
  \begin{threeparttable}
  \resizebox{\textwidth}{!}
  {\begin{tabular}{lp{8cm}l}
    \toprule
    \multicolumn{1}{l}{Alternative} & Alternative attributes & Availability\\
    \midrule
    TRAIN & time, headway, cost (train\_tt, train\_hw, train\_co) & train\_av\\
    SM (Swissmetro) & time, headway, seats\tnote{a}, cost (sm\_tt, sm\_hw, sm\_seats, sm\_co) & sm\_av\\
    CAR & time, cost (car\_tt, car\_co) & car\_av\\
    \toprule
    \multicolumn{1}{l}{Person/Trip variable} & Variable levels\\
    \midrule
    AGE & 0: age $\leq$ 24, 1: 24 $<$ age $\leq$ 30, 2: 39 $<$ age $\leq$ 54, 3: 54 $<$ age $\leq$ 65, 4: 65 $<$ age\\
    MALE & 0: female, 1: male\\ 
    INCOME (thousand CHF per year) & 0: under 50, 1: between 50 and 100, 2: over 100, 3: unknown\\
    FIRST (First class traveler) & 0: no, 1: yes\\
    GA (Swiss annual season ticket) & 0: no GA, 1: owns a GA\\
    PURPOSE & 0: Commuter, 1: Shopping, 2: Business, 3: Leisure\\
    WHO (Who pays) & 0: self, 1: employer, 2: half-half\\
    LUGGAGE & 0: none, 1: one piece, 2: several pieces \\
    \bottomrule
  \end{tabular}}
  \end{threeparttable}
  \begin{tablenotes}
     \footnotesize
     \item[a] a. Seats configuration in Swissmetro: seats=1 if airline seats, 0 otherwise.
  \end{tablenotes}
\end{table}

\subsection{Benchmarks}
We set up four benchmarks: three MNL models and a Mixed Logit model. \textbf{MNL-A} is similar to Bierlaire et al.(2001)'s MNL specification but with some enhancements: 1) the value of travel time and value of headway is made mode-specific; 2) all levels of age and luggage categories are included; and 3) cost coefficients are fixed to -1.0 for directly reading VOT from time coefficients (Table \ref{tab:coef_MNLA}). In the benchmark \textbf{MNL-B}, we add the interaction terms: time*age, time*income and time*purpose (Table \ref{tab:coef_MNLB}). The third benchmark \textbf{MNL-C} is an MNL with all pairs of first-order interactions between characteristics and attributes (Table \ref{tab:coef_MNLC}). This model is equivalent to a TasteNet-MNL with all taste coefficients modeled by a neural network with no hidden layer. The \textbf{Mixed Logit} model has a normally distributed travel time coefficient for each alternative, along with interactions between individual characteristics with travel time for each alternative. Intercepts and coefficients for heading are alternative specific and not randomly distributed. Estimated model coefficients for the benchmarking models are shown in Table \ref{tab:coef_MNLA}, \ref{tab:coef_MNLB},  \ref{tab:coef_MNLC} and \ref{tab:coef_MixedLogit}. 
\begin{table}[!htbp]
 \caption{Estimated Coefficients of MNL-A}
  \centering
  \begin{tabular}{lrrrr}
    \toprule
    Variable Description & Train & Swissmetro & Car\\
    \midrule
    Constant  & 0.1227 & 0.5726 & \\
    Travel time (minutes) & -1.3376 & -1.4011 & -1.0177  \\
    Headway (minutes)    & -0.4509 & -0.8171  \\
    Seats (airline seating = 1) & & 0.1720 \\
    Cost (CHF)        & -1 (fixed) & -1 (fixed) & -1 (fixed) \\
    GA (annual ticket = 1) & 2.0656 & 0.5319 & \\
    Age & & & \\
    \multicolumn{1}{r}{1: 24 $<$ age $\leq$ 30} & -0.7548 & & \\
    \multicolumn{1}{r}{2: 39 $<$ age $\leq$ 54} & -0.9457 & & \\
    \multicolumn{1}{r}{3: 54 $<$ age $\leq$ 65} & -0.4859 & & \\
    \multicolumn{1}{r}{4: 65 $\leq$ age} & 0.6995 & & \\
    Luggage & & & \\
    \multicolumn{1}{r}{1:one piece} & & & -0.1538\\
    \multicolumn{1}{r}{2:several pieces} & & & -0.9230\\
    \bottomrule
  \end{tabular}
  \label{tab:coef_MNLA}
\end{table}
\begin{table}[!htbp]
 \caption{Estimated Coefficients of MNL-B}
  \centering
  
  \begin{tabular}{lrrrr}
    \toprule
    Variable Description & Train & Swissmetro & Car\\
    \midrule
    Constant  & 0.0056 & 0.4674 & \\
    Travel time (minutes) & -0.5006 & -0.4010 & -0.5600  \\
    Travel time * Age & & & \\
     \multicolumn{1}{r}{0: age $\leq$ 24}\\
    \multicolumn{1}{r}{1: 24 $<$ age $\leq$ 39} & -0.6354 & -0.3307 & -0.5696 \\
    \multicolumn{1}{r}{2: 39 $<$ age $\leq$ 54} & -0.8475 & -0.6101 & -0.6105 \\
    \multicolumn{1}{r}{3: 54 $<$ age $\leq$ 65} & -0.1566 & 0.1419 & -0.0915 \\
    \multicolumn{1}{r}{4: 65 $<$ age} & 0.3265 & -0.243 & -0.0234 \\
    Travel time * Income & & & \\
    \multicolumn{1}{r}{0: under 50} & & & \\
    \multicolumn{1}{r}{1: 50 to 100} & -0.2688 & 0.1739 & 0.1623\\
    \multicolumn{1}{r}{2: over 100} & -1.0181 & -0.436 & -0.4093\\
    \multicolumn{1}{r}{3: unknown} & 0.0852 & 0.2828 & -0.0923\\
    Travel time * Purpose & & & \\
    \multicolumn{1}{r}{0: Commute} & & &\\
    \multicolumn{1}{r}{1: Shopping} & -0.2081 & -0.6192 & -0.6062 \\
    \multicolumn{1}{r}{2: Business} & -0.1574 & -0.8688 & -0.1833 \\
    \multicolumn{1}{r}{3: Leisure} & -0.59 & -0.9706 & -0.0162 \\
    
    Headway (minutes)    & -0.6158 & -0.7011  \\
    Seats (airline seating = 1) & & 0.189 \\
    Cost (CHF)        & -1 (fixed) & -1 (fixed) & -1 (fixed)\\
    GA (annual ticket = 1) & 1.6162 & 0.2988 & \\
    
    Luggage & & & \\
    \multicolumn{1}{r}{1:one piece} & & & -0.1714\\
    \multicolumn{1}{r}{2:several pieces} & & & -0.6718\\
    \bottomrule
  \end{tabular}
  \label{tab:coef_MNLB}
\end{table}
\begin{table}[!htbp]
 \caption{Estimated Coefficients of MNL-C}
  \centering
  \resizebox{\textwidth}{!}
  {\begin{tabular}{lrrrrrrrr}
    \toprule
    \multicolumn{9}{c}{Coefficients for alternative attributes} \\
    \cmidrule(r){2-9}
    z (characteristics) & TRAIN\_TT & SM\_TT & CAR\_TT & TRAIN\_HE & SM\_HE & SM\_SEATS & TRAIN\_ASC & SM\_ASC \\
    \midrule
    Intercept & -0.0671 & 0.1455 & 0.0059 & 0.1713 & 0.0646 & 0.3064 & 0.2953 & 0.2067 \\
    Male & -0.1526 & -0.0477 & 0.0742 & -0.2384 & 0.0706 & -0.1016 & 0.0671 & 0.149 \\
    Age \\
    1: (24,39] & -0.0965 & -0.2422 & -0.1093 & 0.0044 & 0.5682 & 0.0517 & -0.1634 & 0.4285 \\
    2: (39,54] & -0.1467 & -0.2022 & -0.195 & -0.2397 & -0.0105 & -0.2135 & -0.2692 & 0.0959 \\
    3: (54,65] & 0.0256 & 0.1201 & 0.0251 & -0.2379 & -0.0807 & 0.1619 & -0.0861 & -0.0344 \\
    4: (65,) & -0.1712 & 0.1435 & 0.1105 & 0.6032 & -0.1488 & -0.1529 & 0.618 & -0.351 \\
    Income \\
    1: 50-100 & 0.0494 & -0.039 & 0.0098 & -0.1884 & -0.2972 & 0.2349 & -0.1776 & 0.1944 \\
    2: over 100 & -0.2825 & -0.1697 & -0.2662 & 0.1393 & 0.0372 & 0.5288 & -0.0406 & -0.0789 \\
    3: unknown & 0.0289 & 0.1467 & -0.2037 & 0.1484 & -0.0721 & -0.4196 & 0.1621 & -0.0459 \\
    First class & -0.1927 & -0.0807 & -0.3297 & -0.4768 & 0.1183 & 0.1302 & 0.2228 & -0.2085 \\
    Who pay \\
    1: employer & -0.2154 & -0.1668 & 0.1231 & 0.028 & -0.0045 & 0.0882 & 0.1191 & 0.3986 \\
    2: half-half & 0.1537 & 0.4771 & 0.4391 & -0.0311 & 0.3917 & 0.3114 & -0.2414 & -0.0332 \\
    Purpose \\
    1:Shopping & 0.2339 & -0.219 & 0.19 & 0.1509 & 0.0493 & 0.1994 & 0.4238 & 0.6996 \\
    2:Business & -0.0872 & -0.3524 & -0.181 & -0.0544 & -0.0195 & -0.0647 & 0.0605 & -0.2941 \\
    3:Leisure & -0.2678 & -0.2778 & -0.0043 & 0.3245 & -0.4552 & -0.0289 & -0.302 & -0.4739 \\
    Luggage \\
    1:one piece & -0.0375 & 0.0861 & 0.2525 & 0.58 & -0.1993 & 0.0413 & 0.3364 & 0.3239 \\
    2:several pieces & 0.022 & -0.1785 & -0.2731 & -0.2946 & 0.0814 & -0.1225 & -0.0041 & 0.2158 \\
    Annual ticket & 0.5912 & -0.0075 & -0.3181 & 0.2652 & -0.2032 & -0.5815 & 0.3576 & 0.1351 \\
    \bottomrule
  \end{tabular}}
  \label{tab:coef_MNLC}
\end{table}
\begin{table}[!htbp]
 \caption{Estimated Coefficients of Mixed Logit Model}
  \centering
  \begin{tabular}{lrrrr}
    \toprule
    Variable Description & Train & Swissmetro & Car\\
    \midrule
    Constant  & & -0.763 & -1.453 \\
    Travel time (minutes) & -1.691 &  -0.265  & -0.897  \\
    Travel time Std (minutes)   & 1.329   &  0.621   &  0.351 \\
    Travel time * Male & -0.537   & -0.149  & -0.009 \\
    Travel time * Age & & & \\
     \multicolumn{1}{r}{0: age $\leq$ 24}\\
    \multicolumn{1}{r}{1: 24 $<$ age $\leq$ 39} & -1.429  & -1.345  & -0.949 \\
    \multicolumn{1}{r}{2: 39 $<$ age $\leq$ 54} & -1.420  & -1.533  & -0.893 \\
    \multicolumn{1}{r}{3: 54 $<$ age $\leq$ 65} & -0.730  & -0.876  & -0.411 \\
    \multicolumn{1}{r}{4: 65 $<$ age} & 0.432 & -0.549  & 0.140 \\
    Travel time * Income & & & \\
    \multicolumn{1}{r}{0: under 50} & & & \\
    \multicolumn{1}{r}{1: 50 to 100} & 0.052  & 0.808  & 0.582 \\
    \multicolumn{1}{r}{2: over 100} & -0.391  & 0.452   & 0.209 \\
    \multicolumn{1}{r}{3: unknown} & 0.899  & 1.359  & 0.633 \\
    Travel time * Purpose & & & \\
    \multicolumn{1}{r}{0: Commute} & & &\\
    \multicolumn{1}{r}{1: Shopping} & -0.754  & -1.676  & -1.312 \\
    \multicolumn{1}{r}{2: Business} & -0.135  & -1.114  & -0.258 \\
    \multicolumn{1}{r}{3: Leisure} & -1.062   & -1.477  & -0.196 \\
    Travel time * Luggage & & & \\
    \multicolumn{1}{r}{1: One piece} & 0.658  & 0.714   & 0.496 \\
    \multicolumn{1}{r}{2: Several pieces}  & -0.922  & -1.477  & -1.597 \\
    Travel time * Annual Ticket & 1.542  & -0.301  & -0.788 \\

    Headway (minutes)    & -0.969 &  -1.365 \\
    Seats (airline seating = 1) & & 0.559 \\
    Cost (CHF)        & -1 (fixed) & -1 (fixed) & -1 (fixed)\\
    \bottomrule
  \end{tabular}
  \label{tab:coef_MixedLogit}
\end{table}

L-MNL is not selected as a benchmark because it is a special case covered by TasteNet-MNL for the Swissmetro dataset: all unused variables in the data-driven part of the L-MNL utility are individual characteristics, so that the data-driven utility in L-MNL corresponds to the ASC taste functions estimated by TasteNet. Given a more complex dataset, we could have a unified version of L-MNL and TasteNet-MNL, whose utility consists of a data-driven component from L-MNL, a TasteNet component, and a knowledge-driven part.

\subsection{Taste-MNL}
\begin{figure}[!htbp]
  \includegraphics[width=\linewidth]{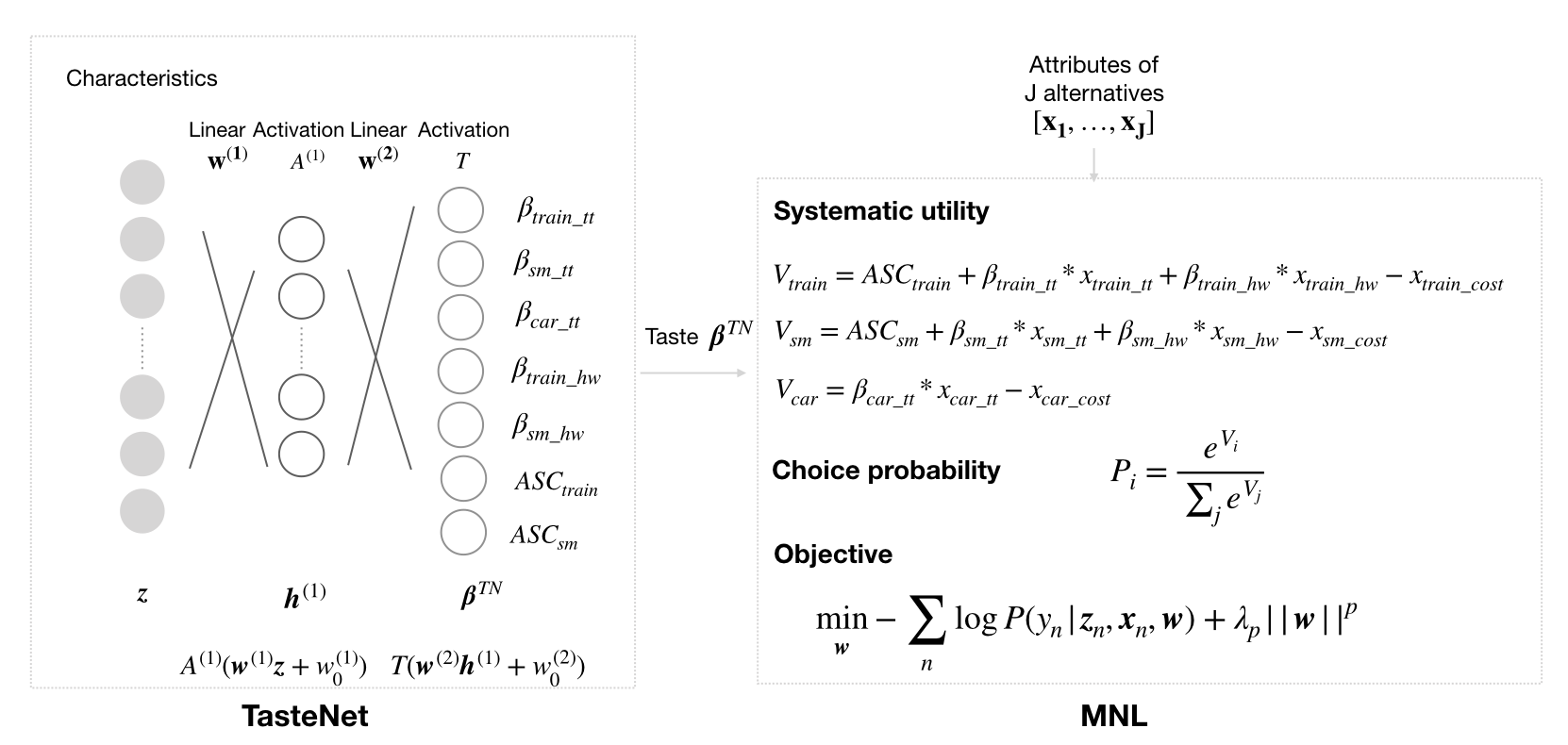}
  \caption{Diagram of TasteNet-MNL for Swissmetro Dataset}
  \label{fig:taste_sm_diag}
  \centering
\end{figure}

The \textbf{TasteNet-MNL} structure for Swissmetro data is shown in Figure \ref{fig:taste_sm_diag}. We specify the utility function of each alternative in the MNL module. Coefficient for cost is fixed to -1 so that the coefficient for time is the negative VOT. There are 7 coefficients in the MNL utilities, including two alternative specific constants. We assume all these coefficients (``tastes") are functions of individual characteristics and model them by TasteNet. This is a special case of the general structure: the set of $\bm{\beta}^{MNL}$ is empty and all taste parameters are modeled by TasteNet as $\bm{\beta}^{TN}$ (Eqn. \ref{eqn:util_tastenet} and \ref{eqn:prob}).  

The TasteNet module consists of a linear layer from input $\bm{z}$ to hidden layer $\bm{h}^{(1)}$, a nonlinear activation $A^{(1)}$ for the hidden layer, followed by a linear layer from hidden layer to output layer and an output activation function $T$ for the output. We choose only 1 hidden layer, since the predicted log-likelihoods on hold-out datasets do not improve with more hidden layers. Input $\bm{z}$ includes all characteristics: age, gender, income, first-class, who pays for travel cost, trip purpose and luggage. We experiment with various sets of hyper-parameters and activation functions shown in Table \ref{tab:taste_sm_hyper}.

Among all TasteNet-MNL scenarios, the one with 110 hidden units, $relu$ for hidden layer activation, negative exponential for non-positive output activation, and no regularization achieves the best prediction performance on the development dataset. 

\begin{table}[!htbp]
     \caption {Options of Hyper-parameters \& Activation Functions}
     \label{tab:taste_sm_hyper}
     \centering
     \begin{threeparttable}
       \begin{tabular}{lc}
       \toprule
         Options & Values \\ \midrule
         Hidden activation & $relu$, $tanh$ \\
         Output activation (for non-positive parameters) & $-relu(-\beta)$, $-exp(-\beta)$ \\
         Hidden layer size & [10, 20, ..., 100] \\
         $l1$ or $l2$ regularization \tnote{a} & [0, 0.0001, 0.001, 0.01] \\
        \bottomrule
       \end{tabular}
       \begin{tablenotes}
         \footnotesize
         \item[a] Either $l1$ or $l2$ regularization is used. 
       \end{tablenotes}
     \end{threeparttable}
   \end{table}

\subsection{Results}

\subsubsection{Prediction Performance}
TasteNet-MNL out-performs all benchmarking models'  prediction accuracy. We use average negative log-likelihood (NLL) and prediction accuracy (ACC) to measure predictability. To avoid model over-fitting, we split data into training, development and test set, and present the performance on each subset. As test set is not used in model training or hyper-parameter selection, prediction performance on it provides a reliable comparison across models. The result is shown in Table \ref{tab:swissmetro_nll}. 

From MNL-A to MNL-C, as more interactions between attributes and individual characteristics are added, the predicted NLL is reduced from 0.755 (MNL-A) to 0.698 (MNL-C). Mixed-Logit's performance is similar to MNL-C (0.703). The best run of TasteNet-MNL (minimal NLL on development set) reduces test NLL from 0.698 to 0.645, and improves ACC from 0.678 to 0.703 and F1 score from 0.574 to 0.62. The mean and standard deviation of each predictability metric from 100 runs also shows consistent improvement: NLL reduced to 0.652, ACC improved to 0.7 and F1 score improved to 0.62. The improved predictability is attributed to the fact that TasteNet learns a more complex representation of systematic taste heterogeneity.  

\begin{table}[!htbp]
 \caption{Average Negative Log-likelihood (NLL), Prediction Accuracy (ACC) and F1 Score by Model}
  \centering
  \resizebox{\textwidth}{!}
  {\begin{tabular}{p{4cm}rrrrrrrrr}
    \toprule
     &\multicolumn{3}{c}{NLL} & \multicolumn{3}{c}{ACC} & \multicolumn{3}{c}{F1}\\
    \cmidrule(r){2-10}
    Model & train & dev & test & train & dev & test & train & dev & test \\
    \midrule
    MNL-A  & 0.762   &  0.728  &  0.755&  0.662   &  0.691  &   0.66 &  0.535 &  0.557 & 0.534\\
    MNL-B  &    0.73   &  0.708  &   0.72   & 0.678   &   0.69  &  0.678   & 0.578 & 0.585 & 0.573 \\
    MNL-C & 0.704	& 0.691	& 0.698 &  0.685	& 0.706	& 0.678 & 0.588 & 0.611 & 0.574 \\
    Mixed-Logit & 0.712 & 0.699 & 0.703 & 0.681 & 0.701 & 0.686 & 0.588 & 0.606 & 0.586 \\
    TasteNet (no2) & 0.607  & 0.646  &  0.645  &  0.737  &  0.718  &  0.703 & 0.668 & 0.634  & 0.620 \\
    TasteNet (100 runs) & 0.605  & 0.650 & 0.652 & 0.735 & 0.718 & 0.703 & 0.674 & 0.645 & 0.621 \\
    mean (std) & (0.0182) & (0.0055) & (0.008) & (0.0094) & (0.0047) & (0.0055) &  (0.016) & (0.009) & (0.010) \\
    \bottomrule
    \end{tabular}}
  \label{tab:swissmetro_nll}
\end{table}

\subsubsection{Estimated Tastes}
Since interpretability is a primary concern for transportation application, we need to investigate whether the interpretations make sense. We show several ways to diagnose. As TasteNet-MNL are direct extensions to MNLs, and Mixed Logit does not show superior predictability, we focus on the comparisons between MNLs and TasteNet-MNL's interpretations here and in the following sections. 

First, we compare the average taste estimations for observations in the Swissmetro dataset by different models. We apply each model to obtain individual estimates, and compute the averages (Table \ref{tab:sm_mean_taste}). Since the cost coefficients are fixed to -1.0, all taste coefficients are in the willingness-to-pay (WTP) space measured by Swiss Franc (CHF). 

\begin{table}[!htbp]
 \caption{Average Tastes Estimated by Different Models}
 \label{tab:sm_mean_taste}
  \centering
  \resizebox{\textwidth}{!}
  {\begin{tabular}{lcrrrrr}
  \toprule
  Mode & Taste & MNL-A	& MNL-B	& MNL-C	& TasetNet (vs MNL-C) & TasteNet mean (std)\\
  \midrule
    TRAIN	& TT	& -1.338	& -1.710	& -1.846	& -2.327 (26\%) & -2.376 (0.179)\\
	& HE	& -0.451	& -0.616	& -0.880	& -1.102 (25\%) & -1.161 (0.077)\\
	& ASC	& -0.198	& 0.234		& 0.368		& 0.801 (117\%) & 0.824 (0.135) \\
    SM	& TT	& -1.401	& -1.514	& -1.505	& -1.764 (17\%) & -1.905 (0.135)\\
	& HE	& -0.817	& -0.701	& -1.039	& -1.733 (67\%) & -1.709 (0.131) \\
	& SEATS	& 0.172		& 0.189		& 0.420		& 0.266 (-37\%) & 0.270 (0.079) \\
	& ASC	& 0.648		& 0.510		& 0.512		& 0.669 (31\%) & 0.675 (0.064) \\
    CAR	& TT	& -1.018	& -1.251	& -1.354	& -1.685 (24\%) & -1.713 (0.101) \\
    \bottomrule
  \end{tabular}
  }
\begin{tablenotes}
 \footnotesize
 \item TT: time. HE: headway. ASC: alternative specific constant
\end{tablenotes}
\end{table}

We find that from MNL-A to MNL-C, average values of travel time (VOT) and values of headway (VOHE) increase with more interaction terms being added to the utility function. For example, train VOT increases from 1.34 to 1.85 CHF per minute. Swissmetro VOT increases from 1.4 to 1.51 CHF per minute, and car VOT rises from 1 to 1.35 CHF per minute. Both MNL-B and MNL-C suggest that VOT of train is higher than that of Swissmetro or car.  MNL-C also gives higher average VOHE for both train and Swissmetro than MNL-B or MNL-A.

TasteNet-MNL gives the largest average VOT and VOHE among all models (Table \ref{tab:sm_mean_taste}). Its average VOT estimates for train, swissmetro and car are 26\%, 17\%, and 24\% higher, respectively than those predicted by MNL-C. Its average VOHE estimates for train and swissmetro are 25\% and 67\% higher than those estimated by MNL-C. 

\begin{sidewaysfigure}[!htbp]
    \includegraphics[width=\textwidth]{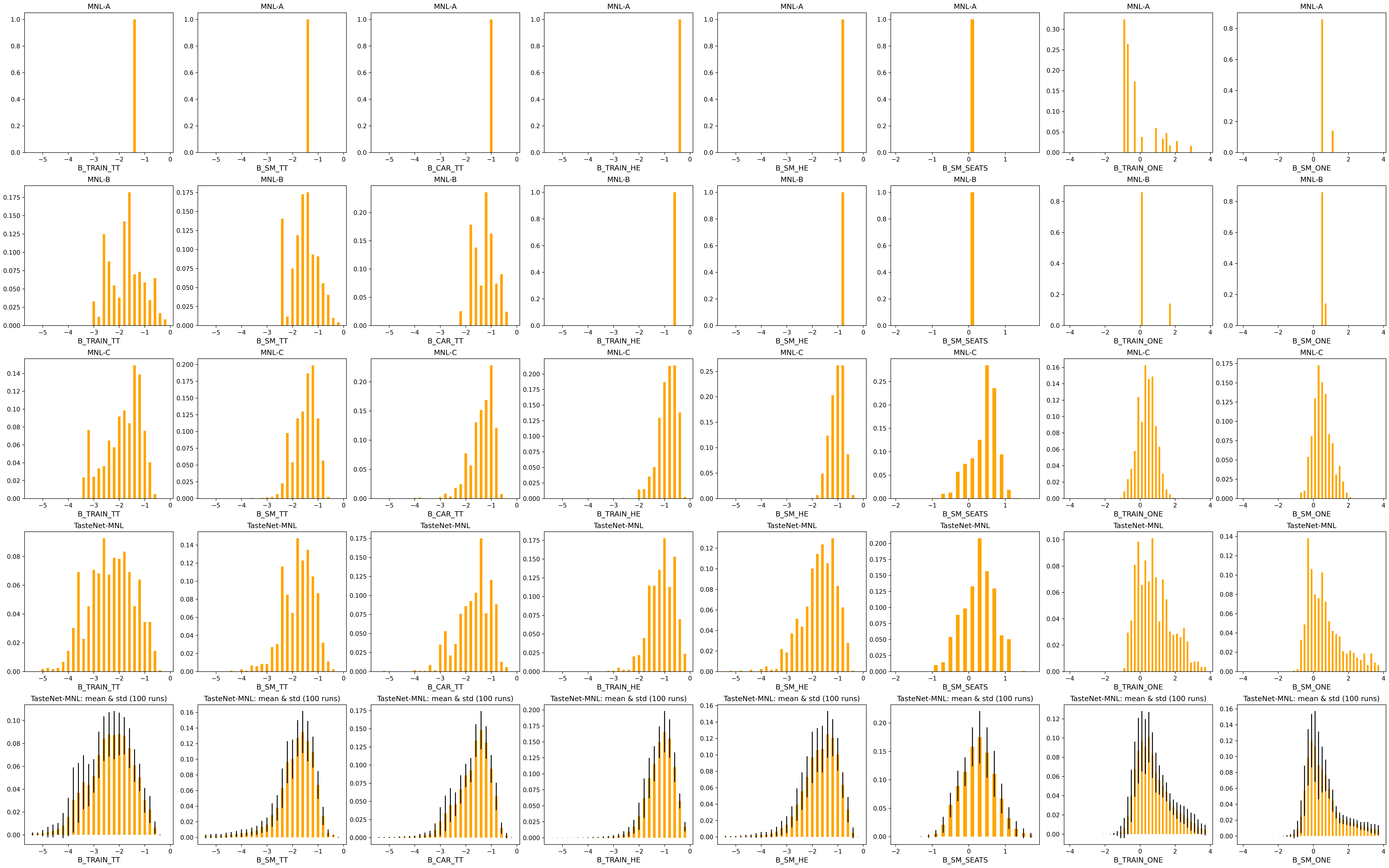}
    \caption{Population Taste Distributions by Models
  (Swissmetro Dataset)}
    \label{fig:sm_hist_taste}
\end{sidewaysfigure}

We further investigate where the higher average VOTs come from. We plot histograms for each taste parameter and for each model (Figure \ref{fig:sm_hist_taste}). As interactions are incrementally added from MNL-A to MNL-C, the model captures more taste variations. Compared to MNLs with linear utilities, TasteNet-MNL discovers a wider range of taste variations. In particular, the VOTs and VOHEs for all travel modes have longer tails on the high end of WTP. The last row of histograms (Figure \ref{fig:sm_hist_taste}) shows the mean and standard deviation from 100 runs. Overall, we see a consistent pattern in the distributions, with different amount of variability across estimation runs. There is greater uncertainty in Train's VOT estimates, and Train and Swissmetro's ASC estimates. The bigger uncertainty in Train's VOT (compared to Car or Swissmetro's) coincides with the larger standard deviation of the randomly distributed travel time coefficient for Train (1.33), compared to the standard deviation of 0.62 for Swissmetro and 0.35  for Car (Table \ref{tab:coef_MixedLogit}).

Based on the synthetic data experiments, we have confidence to believe that TasteNet-MNL's better predictability is due to more accurately capturing the underlying taste functions.  However, readers should be aware of the risk of over-fitting TasteNet to outliers, especially when dealing with a  noisier dataset or using more complex neural networks. We recommend conducting a thorough interpretability check at both the aggregate and the disaggregate level, and comparing the results against the benchmarking models to spot any spurious behaviors. The next two sections provide additional ways to check. 

\subsubsection{Taste Functions}
An interpretable model should provide a reasonable function relationship between choice outcome and input at the disaggregate level. We propose a diagnostic tool for model interpretability: visualizing the taste function.

Each model provides a taste function that maps individual characteristics to a type of taste value (e.g. VOT, VOHE).  We want to check whether TasteNet-MNL learns sensible taste functions at the individual level, in comparison to benchmarking MNLs. 

Since function input $\bm{z}$ is multi-dimensional, we cannot directly visualize the functions. Instead, we pick an individual with characteristics $\bm{z}$. We vary one dimension of $\bm{z}$: $z_i$,  while keeping other dimensions fixed $z_{j\neq i}$. We plot a particular taste parameter as a function of $z_i$: $\beta_k = f_{model}(z_i; z_{j\neq i})$. 

\begin{figure}[!htbp]
  \includegraphics[width=\linewidth]{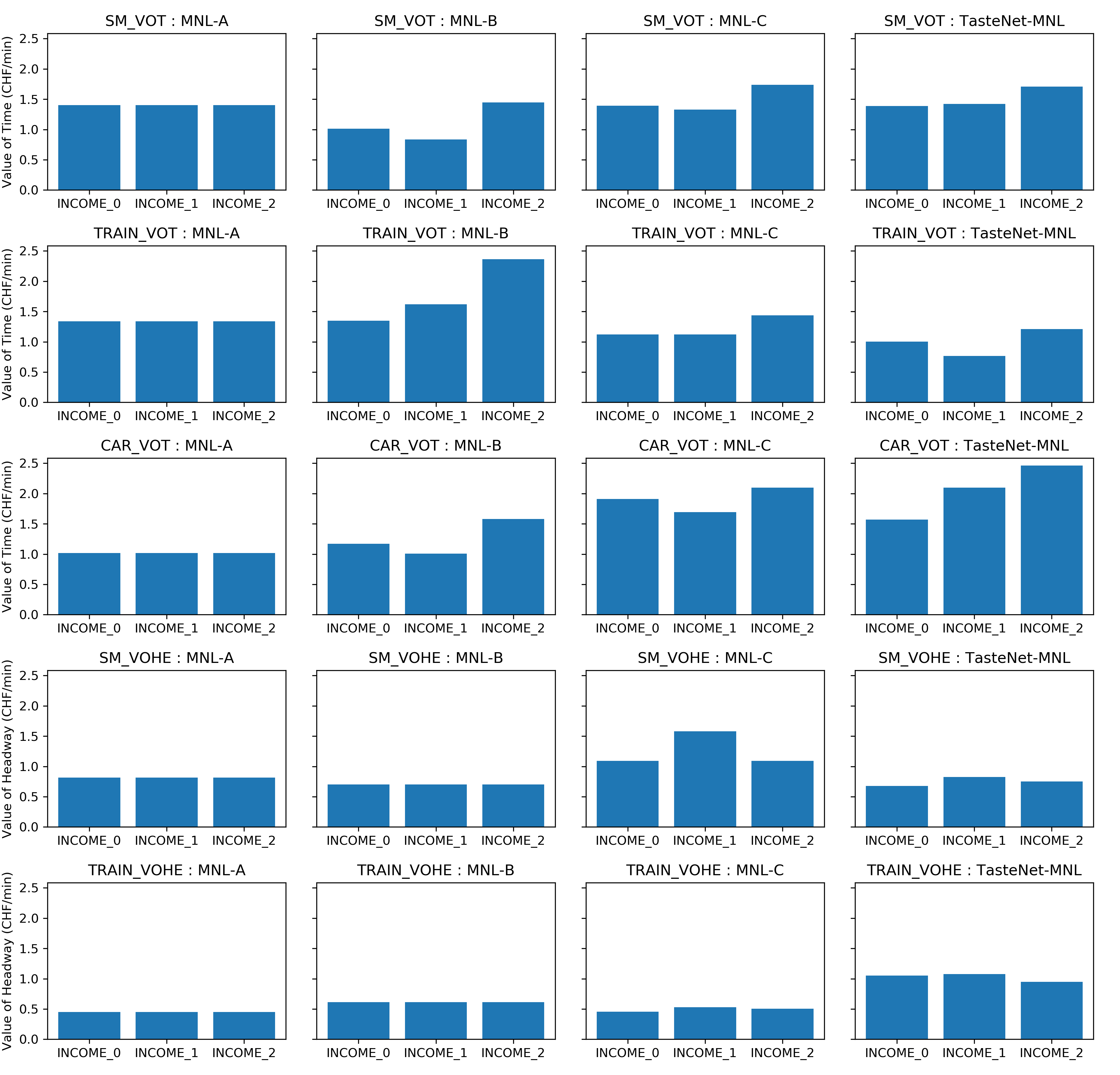}
  \caption{Tastes as Functions of Income for a Selected Person by Different Models}
  \label{fig:taste_func}
  \centering
\end{figure}

For example, we pick a person with characteristics shown in Table \ref{tab:sm_example_person}. We vary this person's income and ask each model a question: what are the VOTs for such a person as his income varies? We compare the answers given by different models. Figure \ref{fig:taste_func} shows the VOTs and VOHEs estimated by different models versus income with other characteristics fixed. 

\begin{table}[!htbp]
    \centering
    \caption{Example Person Selected for Comparing Taste Function by Models}
    \label{tab:sm_example_person}
    \begin{tabular}{lll}
        \toprule
        & Characteristics & Value\\
        \midrule
        $\bm{z}_{fixed}$ & MALE & Male  \\
        & AGE & (39,54]\\
        & PURPOSE & Commute\\
        & WHO & Self\\
        & LUGGAGE & One piece\\
        & GA & Yes\\
        & FIRST & No\\
        \midrule
        $\bm{z}_{vary}$ & INCOME & 
        {0: under 50, 1: 50 to 100, 2: over 100}\\
         \bottomrule
    \end{tabular}
\end{table}

Compared with the benchmarking MNLs, VOT and VOHE estimated by TasteNet-MNL all fall within credible ranges. TasteNet-MNL gives more or less different estimates. Swissmetro VOT estimates are not very different between TasteNet-MNL and MNL-C. Regarding train VOT, TasteNet-MNL gives smaller estimates for all three income groups than MNL-C. Car VOT estimated by TasteNet-MNL is higher for higher-income groups and lower for the lowest income group. With respect to VOHEs, TasteNet-MNL gives higher estimates for train and lower estimates for Swissmetro for all income levels. MNL-C shows a monotonic relationship between VOT and income only for train VOT, while TasteNet-MNL identifies the monotonicity for swissemtro VOT and car VOT. 

As we do not know the ground truth, the interpretability and credibility of the model inevitably depend on expert knowledge and judgment. We draw many individual cases,  visualize the taste functions, and compare across models. Overall, taste parameters by TasteNet-MNL fall within similar ranges as the MNLs do. Yet a particular taste of a specific individual given by TasteNet-MNL can agree with or differ from the MNLs. Based on TasteNet-MNL's better predictability, we trust that it obtains more accurate taste estimates for individuals. 

\subsubsection{Elasticity}
Elasticity is another aspect to check model interpretability and catch abnormal neural network behaviors. We first apply each model to calculate disaggregate point elasticity of Swissmetro mode choice with respect to Swissmetro travel time for each observation. The point elasticity equation for MNL models is shown in Eqn. \ref{eqn:elas_mnl} and for TasteNet\_MNL in Eqn. \ref{eqn:elas_TN}. TasteNet-MNL's individual elasticity estimates differ from MNL-C by 0.2287 on average. 

With individual elasticities, we then compute aggregate elasticity, which measures a group of decision-makers' response to an incremental change in a variable. This is defined in Eqn \ref{eqn:aggregate_elas} as the percentage change in the expected share of the group choosing alternative $i$ ($W_i$) with respect to one percentage change in variable $x_{ki}$. It is equivalent to a weighted average of the individual elasticities using the choice probabilities as weights.

\begin{equation}
    E_{x_{kin}}^{P_n(i)}=(1-P_n(i))x_{kin}\beta_k
    \label{eqn:elas_mnl}
\end{equation}

\begin{equation}
    E_{x_{kin}}^{P_n(i)}=(1-P_n(i))x_{kin}\beta_k(\bm{z}_n)
    \label{eqn:elas_TN}
\end{equation}

\begin{equation}
    E^{W(i)}_{x_{ki}} = \frac{\partial W(i)}{\partial x_{ki}}\frac{x_{ki}}{W(i)} = \frac{\sum_n P_n(i)E_{x_{kin}}^{P_n(i)}}{\sum_n P_n(i)}
\label{eqn:aggregate_elas}
\end{equation}

The aggregate elasticities of Swissmetro mode share with respect to Swissmetro travel time are -0.43, -0.45, and -0.41 for MNL-A, MNL-B and MNL-C, compared to -0.437 for TasteNet-MNL. We further compare aggregate elasticity by group, such as income (Table \ref{tab:agg_elas_income}). TasteNet-MNL suggests higher elasticities for low income and high income groups than MNL-C. But overall, TasteNet-MNL gives choice elasticities close to MNLs and within reasonable range. We recommend performing this experiment systematically against a dummy model (e.g. MNL-C) as a sanity check. 

\begin{table}[!htbp]
    \centering
    \caption{Aggregate Choice Elasticity of Swissmetro w.r.t Travel Time by Income Group}
    \label{tab:agg_elas_income}
    \begin{tabular}{lrrrr}
    \toprule
        &	INCOME0	& INCOME1 &	INCOME2\\
    \midrule
        MNL-A &	-0.3765 &	-0.4297	&   -0.4706\\
        MNL-B &	-0.3975 &	-0.3923 &	-0.5329\\
        MNL-C &	-0.3759	&   -0.3706	&   -0.4653\\
        TasteNet-MNL &	-0.4200	& -0.3982 &	-0.4810\\
    \bottomrule
    \end{tabular}
\end{table}
\section{Conclusions \& Discussions}
In this paper, we utilize a neural network to learn taste representation while keeping model interpretability. Departing from a traditional either-or approach, we integrate neural networks and DCMs to take advantage of both. 

With a synthetic dataset, we show that TasteNet-MNL can learn the underlying taste function. It achieves the same level of accuracy as the true model in predicting choice and deriving economic indicators. Exemplary MNLs with misspecified utility result in parameter bias and lower prediction accuracy. On the Swissmetro dataset, TasteNet-MNL not only surpasses MNLs and Mixed Logit model benchmarks in predictability, but also achieves interpretable economic indicators: individual-level VOTs and elasticities from TasteNet-MNL are comparable to the results of the benchmarking MNLs. Moreover, TasteNet-MNL discovers a greater range of taste variations in the population than the benchmarking MNLs. The average VOT estimates by TasteNet-MNL are higher than the MNLs'. 

We demonstrate the benefit of integrating neural networks and DCM. Neural networks can learn complex functions from data and reduce bias in manual specifications. The theory behind a DCM can guide the neural networks to output meaningful results. A high-level idea behind TasteNet-MNL is to assign the more complex or unknown part of the model (e.g. taste heterogeneity) to a neural network, and keep the well-understood part (e.g. the trade-offs between alternative attributes) parametric. This idea can be applied to other settings. For example, a neural network can be used to model class membership in a latent class choice model.

TasteNet-MNL is most suitable for a fairly large dataset (thousands of examples or more) with a rich set of individual characteristics. Modelers can compare TasteNet-MNL against MNL benchmarks provided by human experts to understand the predictability gap and existence of misspecified taste heterogeneity. Modelers should apply proper regularization strategies in order not to over-fit the model. Based on our findings from the synthetic data experiments, TasteNet-MNL yields large estimation errors on out-of-distribution samples. Users should be aware of this risk and careful when applying the model to samples near or outside the training data's input distribution. We recommend estimating the models for multiple times (e.g. 100) with different initialization to measure the uncertainty in estimates; and conducting a systematic sanity check of the behavioral indicators (e.g. value of time, elasticity) against baseline models to spot spurious behaviors.


There are limitations with our approach and open questions for future research. First, the TasteNet-MNL model only accommodates systematic taste variations. \textit{Random taste heterogeneity} is an important source of heterogeneity. How to model distributions of taste parameters with neural networks is an intriguing question for future research. \citet{han2019} proposes a neural network embedded latent class choice model, as one way to represent random heterogeneity. Future work can develop a neural embedded continuous mixed logit model.

Second, TasteNet-MNL focuses on modeling \textit{taste} heterogeneity. Non-linear effects of \textit{attributes} have been observed empirically \citep{Monroe1973, gupta1992discounting, kalyanaram1994}. Nonlinear effects, such as the saturation effect and threshold effect, are explained by prospect theory and assimilation-contrast theory \citep{kahnemann1979prospect, winer1986, winer1988}. Future work may extend TasteNet-MNL model to reflect nonlinearity in attributes.

Lastly, we suggest comparing TasteNet-MNL with DCMs under various empirical settings, in terms of prediction performance and behavioral interpretations. We find in the Swissmetro case study that TasteNet-MNL yields higher average VOTs and a greater variety of tastes than MNLs. Future research may test if this finding holds for other datasets, and understand the practical implications for demand forecast and scenario analysis.
\clearpage

\textbf{Acknowledgements}
This work was supported by the New England University Transportation Center at MIT [grant number: DTRT13-G-UTC31; funding agency: USDOT/RITA.]

\bibliography{main}
\clearpage
\appendix

\section{Synthetic Data Generation}
\label{app-A}

We first draw input characteristics $\bm{z}$ according to the input distribution described in Table \ref{tab:taste_toy_desc}. Alternative attributes $cost$, $time$ and $wait$ are drawn from the ranges described in Table \ref{tab:taste_toy_desc}. For the TOY\_UNCORREL dataset, $time$ and $wait$ are drawn from independent uniform distributions. For the TOY\_CORREL dataset, $time$ and $wait$ are drawn from a bivariate uniform distribution with a correlation of 0.6.

With the true model, we compute choice probabilities for each observation. Finally, we draw a chosen alternative for each observation according to the predicted choice probabilities. We generate 10000, 2000 and 2000 examples for training, development and test data, respectively. Training data is used for model estimation. The development set is for selecting hyper-parameters. The test set is not used in training or selection. It evaluates models' generalization ability. 
\begin{table}[!htbp]
 \caption{Description of Explanatory Variables for Synthetic Datasets}
  \centering
  \resizebox{\textwidth}{!}
  {\begin{tabular}{llllp{5cm}}
    \toprule
    & & Variable & Description & Distribution\\
    \hline
    Individual characteristics & $z_1$  & inc & Income (\$ per minute) & {LogNormal(log(0.5),0.25) for full-time; \newline LogNormal(log(0.25),0.2) for not full-time}\\
    & $z_2$ & full & Full-time worker (1=yes, 0=no) & Bern(0.5)\\
    & $z_3$ & flex & Flexible schedule (1=yes, 0=no) & Bern(0.5)\\
    \midrule
    Attributes (for TOY\_UNCORREL) & $x_1$ & cost & Cost (\$) & 0.2 to 100\$ \\
    (uncorrelated time and wait) & $x_2$ & time & Travel time (minutes) & 5 to 100 minutes\\
    & $x_3$ & wait & Waiting time (minutes) & 5 to 30 minutes\\
    \midrule
    Attributes (for TOY\_CORREL) & $x_1$ & cost & Cost (\$) & 0.2 to  100\$ \\
   (correlated time and wait, & $x_2$ & time & Travel time (minutes) & 5 to 100 minutes\\
    correlation = 0.6) & $x_3$ & wait & Waiting time (minutes) & 5 to 30 minutes\\
    \bottomrule\end{tabular}}
  \label{tab:taste_toy_desc}
\end{table}

\end{document}